\documentclass[12pt,letterpaper]{article}
\usepackage{jheppub}

\usepackage{graphicx}
\usepackage{bbm}
\usepackage{amsmath}
\usepackage{amssymb}
\usepackage{mathtools}
\usepackage{tocvsec2}

\usepackage[labelformat=simple]{subcaption}

\newcommand{\cA}{\mathcal{A}}

\newcommand{\cG}{\mathcal{G}}

\newcommand{\cL}{\mathcal{L}}
\newcommand{\cM}{\mathcal{M}}

\newcommand{\cP}{\mathcal{P}}
\newcommand{\cQ}{\mathcal{Q}}

\newcommand{\cT}{\mathcal{T}}

\newcommand{\df}{\mathrel{:=}}
\newcommand{\noeq}{\mathrel{\phantom{=}}}

\newcommand{\f}{\mathfrak{f}}

\newcommand{\SO}[1]{\mathrm{SO}\!\left(#1\right)}

\newcommand{\Sp}[1]{\mathrm{Sp}\!\left(#1\right)}
\newcommand{\U}[1]{\mathrm{U}\!\left(#1\right)}

\newcommand{\ISO}[1]{\mathrm{ISO}\!\left(#1\right)}
\newcommand{\gO}[1]{\mathrm{O}\!\left(#1\right)}

\DeclareMathOperator{\diag}{diag}

\newcommand{\II}{\ensuremath{I\!I}}

\newcommand{\secref}[1]{\S\ref{#1}}

\begin{document}
\begin{titlepage}

\setcounter{page}{1} \baselineskip=15.5pt \thispagestyle{empty}

{\flushright ACFI-T20-07 \\ }

\bigskip\

\vspace{1.05cm}
\begin{center}
{\LARGE \bfseries Black Holes, Moduli, \\ \vspace{0.3cm} and Long-Range Forces}

 \end{center}
\vspace{1cm}

\begin{center}
\scalebox{0.95}[0.95]{{\fontsize{14}{30}\selectfont Ben Heidenreich$^{a}$}}
\end{center}

\begin{center}
\vspace{0.25 cm}
\textsl{$^{a}$Department of Physics, University of Massachusetts, Amherst, MA 01003 USA}\\

\vspace{0.25cm}
\end{center}

\vspace{1cm}
\noindent

It is well known that an identical pair of extremal Reissner-Nordstr\"om black holes placed a large distance apart will exert no force on each other. In this paper, I establish that the same result holds in a very large class of two-derivative effective theories containing an arbitrary number of gauge fields and moduli, where the appropriate analog of an extremal Reissner-Nordstr\"om black hole is a charged, spherically symmetric black hole with vanishing surface gravity or vanishing horizon area. Analogous results hold for black branes.

\vspace{1.1cm}

\bigskip
\noindent\today

\end{titlepage}

\maxtocdepth{subsection}
\maxsecnumdepth{subsection}

\hrule
\tableofcontents

\bigskip\medskip
\hrule
\bigskip\bigskip

\section{Introduction}

Massless scalar fields with exactly vanishing potentials---i.e., moduli---are ubiquitous in string-derived quantum gravities with unbroken supersymmetry. When present, moduli have important low-energy consequences, mediating new long-range forces in addition to the usual gauge and gravitational ones.

These forces are often required to fulfill the predictions of supersymmetry. For instance, mutually supersymmetric (BPS) objects must have vanishing force between them, but the electrostatic and gravitational forces do not cancel in general. Moduli-mediated forces exactly make up the difference, bringing the net force to zero. Likewise, supersymmetric black holes typically have more charge than is allowed by the Reissner-Nordstr\"om (RN) extremality bound $M^2 \ge \gamma_{\rm RN} Q^2$. The discrepancy is again explained by the moduli, which alter the external geometry of the black hole and reduce the gauge coupling near its core, avoiding a naked singularity.

In the absence of moduli, there is a simple connection between long-range forces and black holes: an extremal Reissner-Nordstr\"om (RN) black hole carries just enough charge so that two identical copies of the black hole have vanishing long-range force between them (the extremal black hole has vanishing ``self-force''). In other words, extremal RN black holes behave like BPS objects, regardless of whether they preserve any supersymmetry. This holds to leading order in the derivative expansion of the effective field theory---hence, for parametrically large black holes---but not necessarily beyond that. 

\bigskip

In this paper, I establish that the connection between extremality and vanishing self-force persists in a general theory of moduli, Abelian gauge fields, and Einstein gravity at the two-derivative level with vanishing cosmological constant, for an appropriate generalization of ``extremal.'' 

In particular, an extremal RN black hole has vanishing Hawking temperature. In theories with a dilaton, ``extremal'' black holes do not always have this property~\cite{gibbons:1987ps}, see also~\secref{subsec:dilBHs}, but instead the Bekenstein-Hawking entropy goes to zero in the extremal limit. 
Let us label a black hole with either vanishing Hawking temperature (surface gravity) or vanishing Bekenstein-Hawking entropy (horizon area) \emph{quasiextremal}. 

I will show that static spherically symmetric quasiextremal black holes have vanishing long-range self-force, whereas all other \emph{non-extremal} (finite temperature and entropy) static spherically symmetric black holes are self-attractive. As before, these results hold at the two derivative level, hence for parametrically large black holes, but not necessarily beyond that. Similar results will be obtained for black branes.

\bigskip

The connection between quasiextremality and long range forces is particularly relevant for the Repulsive Force Conjecture~\cite{Palti:2017elp,Lust:2017wrl,SimonsTalk,Lee:2018spm,Heidenreich:2019zkl} variant of the Weak Gravity Conjecture~\cite{Arkanihamed:2006dz}.
A different notion of ``extremal'' is kinematically relevant. Let us label a black hole that is lighter than all others of the same charge \emph{extremal}. BPS black holes are always extremal because of saturating a BPS bound. More generally, extremal black holes are closely related to the Weak Gravity Conjecture when formulated kinematically as in, e.g.,~\cite{cheung:2014vva, Heidenreich:2015nta}. 

It is relatively easy to write down effective field theories admitting quasiextremal black hole solutions that are not extremal. For instance, if the gauge coupling has a critical point somewhere in moduli space then there are RN black hole solutions in this vacuum, including one which is quasiextremal (zero temperature), but if the critical point is not a local minimum then it can be shown (e.g., using the methods of~\cite{ExtPaper}) that there are lighter black hole solutions of the same charge.

On the other hand, an extremal black hole should be quasiextremal on physical grounds, because if not it will emit finite temperature Hawking radiation from a horizon of finite area, and thus the radiated power will be nonzero and the black hole will lose mass over time, contradicting the assumption that it was the lightest possible black hole of that given charge. This argument relies on black hole quantum mechanics and assumes that finite-temperature Hawking radiation does not efficiently discharge the black hole, but a purely classical proof that extremal black holes are quasiextremal can be developed~\cite{ExtPaper}.

The relationship between quasiextremal and extremal black holes is particularly important for understanding the relationship between the Repulsive Force Conjecture and the Weak Gravity Conjecture~\cite{Heidenreich:2019zkl}.\footnote{Other connections between these two conjectures have been proposed~\cite{Lee:2018spm,Gendler:2020dfp}, simultaneously incorporating the Swampland Distance Conjecture~\cite{Ooguri:2006in}.} This will be explored in more detail in a companion paper~\cite{ExtPaper}, where a general prescription for determining the mass of an extremal black hole and hence the \emph{extremality bound} $M_{\text{BH}}(Q) \ge M_{\text{ext}}(Q)$ will be discussed. In particular, this prescription, based on~\cite{Ceresole:2007wx,Andrianopoli:2007gt}, is intimately tied to the BPS-like no-force condition for quasiextremal black holes derived in this paper, and will lead to a number of useful theorems relating extremality and self-force.

\bigskip

To obtain my primary result, I will show that static spherically symmetric black hole solutions and static spherically symmetric and worldvolume translation invariant black brane solutions are determined by a set of equations with a universal form at two-derivative order in the derivative expansion. Among these equations is a first order (constraint) equation descending from the Einstein equations that fixes the self-force in terms of the product of the surface gravity and the horizon area of the solution.

To derive these equations, I make some very general assumptions about the form of the two-derivative effective action for scalars, $p$-form gauge fields, and gravity. These assumptions are difficult to derive from first principles (due to the possibility that the gauge symmetry may take an unusual form), but are consistent with most if not all examples that arise in UV complete quantum gravities such as string theory.

In addition, in this paper I only consider black hole solutions that do not cross from one branch of the moduli space to another. The fact that such a crossing \emph{can} occur does not seem to be generally known, and the details are sufficiently interesting to warrant a dedicated treatment, see~\cite{CornersPaper}. In the end, the self-force properties of these novel solutions are the same as those discussed in this paper.

\bigskip

An outline of this paper is as follows. Quasiextremal black holes and black branes are introduced in~\secref{sec:quasiextremal}.
In~\secref{sec:setup}, I discuss the low energy effective action and isolate those terms which are relevant to the following analysis.
Long range forces between particles and parallel branes are discussed in~\secref{sec:forces}. In~\secref{sec:BH}, I discuss black hole and black brane solutions and their thermodynamics and show that quasiextremal black holes and black branes have vanishing self-force, whereas non-extremal black holes and black branes are self-attractive. Appendix~\ref{app:democratic} treats theta angles, magnetic charges, and self-dual gauge fields in various dimensions. Appendix~\ref{app:examples} reviews some examples from the literature.

\section{Quasiextremal black holes and black branes} \label{subsec:extremalityDefn} \label{sec:quasiextremal}

For the purposes of this paper, a black hole is a smooth asymptotically flat\footnote{A black hole could also be asymptotically AdS or asymptotically dS, or perhaps have some more exotic asymptotics, but this is not relevant for the present paper.} solution to a gravitational theory with some mass $M$, charge $Q$, and angular momentum $L$, such that the source of the mass, charge, and angular momentum is hidden behind a single smooth event horizon with a spherical topology. Note that $Q$ and $L$ are vectors whose dimensions depend on the gauge group and the spacetime dimension, respectively. For the remainder of the paper, I will only consider static and spherically symmetric black holes, so $L$ is necessarily vanishing.

The thermodynamic properties of black holes are closely related to their horizon area $A \ge 0$ and surface gravity $\kappa \ge 0$. The former determines the Bekenstein-Hawking entropy, and the latter determines the Hawking temperature.

Black holes with both $\kappa > 0$ and $A > 0$ are \emph{non-extremal}. Such black holes have finite temperature and finite entropy, and behave thermodynamically. By comparison, I will call black holes with either $\kappa = 0$ or $A=0$ \emph{quasiextremal}.\footnote{Solutions with $A=0$ are commonly known as ``small black holes.''} In part because a smooth event horizon cannot have vanishing area, it is convenient to include in the latter class solutions with a singular horizon that are the limit of some sequence of solutions with a smooth horizon. Despite the singularity, these solutions can be physically important. For instance, D0 brane solutions in string theory are of this type.

Thermodynamically, $\kappa = 0$ solutions have vanishing Hawking temperature, hence they do not emit Hawking radiation. Likewise, a system with vanishing entropy cannot give off heat, and so $A=0$ solutions also do not emit Hawking radiation.\footnote{Heuristically, this is because the emitting surface shrinks to zero size; correctly understanding the details of this mechanism is beyond the scope of this paper.} Thus, unlike non-extremal black holes, quasiextremal black holes do not radiate and are metastable. 

\bigskip

A black $(p-1)$-brane is an extended object with $p$ worldvolume spacetime dimensions. For a black brane with an infinite planar worldvolume, the horizon topology $S^{d-p-1} \times \mathbb{R}^{p-1}$ is required and (for vanishing cosmological constant) the solution must be asymptotically flat far from the brane worldvolume.\footnote{This is a distinct and weaker requirement than simply ``asymptotically flat'' because the brane worldvolume stretches off to infinity.} Black $(p-1)$-branes can be charged under $p$-form gauge fields $A_p$ and can carry angular momentum density in their transverse directions. For the remainder of this paper, I will only consider static, spherically symmetry, uniform and isotropic\footnote{That is, solutions must be invariant under rotations tranverse to the brane (spherically symmetric) as well as spatial translations (uniform) and rotations (isotropic) along the brane worldvolume.} black branes, so in particular their angular momentum density vanishes.

The horizon area of such a black brane is typically infinite, but the horizon area per unit spatial worldvolume $\cA$ is finite. Thermodynamically, this determines the entropy density of the brane. Similar to before, a black brane with $\kappa > 0$ and $\cA > 0$ is \emph{non-extremal}, whereas one with either $\kappa = 0$ or $\cA = 0$ is \emph{quasiextremal}. As before, the quasiextremal class is taken to include solutions with a singular horizon that are the limit of some sequence of solutions with smooth horizons. In fact, it turns out that all quasiextremal black branes are boost-invariant along their worldvolumes, implying a singular horizon. This includes all BPS brane solutions in string theory.

By the same reasoning as before, quasiextremal branes do not give off Hawking radiation, whereas non-extremal branes do.\footnote{Moreover, quasiextremal branes seem to be immune to the Gregory-Laflamme instability~\cite{Gregory:1994tw} that afflicts non-extremal black branes~\cite{Gregory:1993vy,Gregory:1994bj}.}

\section{The two-derivative effective action} \label{sec:setup}

To study black holes and their long range forces, I assume that low-energy, long-wavelength physics is described by a weakly-coupled effective action, organized in a derivative expansion. For parametrically large black holes, only the leading, two-derivative effective action will be important, and I focus on this exclusively for the rest of the paper. In the same limit, any massive fields can be integrated out. This generates higher-derivative corrections, but these can be ignored for parametrically large black holes as before, so that in the end we obtain a two-derivative effective action for the massless fields only.

At tree-level massless fermions neither affect black hole solutions nor mediate long-range forces, so we ignore them for the time being. By well-known arguments, massless fields can have spin at most 2,\footnote{I assume that the number of massless fields is finite, thereby excluding exotic possibilities such as Vasiliev theories.} and so the bosonic effective action depends only on the metric (spin 2), $p$-form gauge fields (spin 1), and scalar fields (spin 0).

The typical structure of this effective action is as follows: for each $p$-form gauge field $A_p$, there is a gauge-invariant modified field strength $\tilde{F}_{p+1} = d A_p + (\ldots)$ from which the kinetic term is built, where the omitted terms involve wedge products of other $q$-forms and their exterior derivatives. In general, gauge transformations on these $q$-forms $B_q \to B_q + d \lambda_{q-1}$ do not leave $d A_p$ invariant, but the extra terms in the modified field strength $\tilde{F}_{p+1}$ ensure that it is gauge invariant.

We can constrain these extra terms by considering the modified Bianchi identity:
\begin{equation}
d \tilde{F}_{p+1}^a = \alpha^a_{\; b} \tilde{F}_{p+2}^b + \beta^a_{\; b} \wedge \tilde{F}_{p+1}^b + \gamma^a_{\; b} \wedge \tilde{F}_p^b + \ldots \,,
\end{equation}
where we isolate the $q$-form of the highest rank in each term, $a,b,\ldots$ are indices labelling the different forms fields and $\alpha^a_{\; b}$, $\beta^a_{\; b}$, and $\gamma^a_{\; b}$ are zero, one, and two-forms respectively built from the other fields.

The first term comes from $\tilde{F}_{p+1} = d A_p+ \alpha A_{p+1}+(\ldots)$, where the indices $a,b,\ldots$ are temporarily suppressed for simplicity. However, this generates a Stueckelberg mass for $A_{p+1}$ (eating $A_p$, which can be set to zero by a gauge transformation on $A_{p+1}$). Since we have already integrated out all massive fields, such a coupling cannot occur, hence $\alpha^a_{\; b}=0$.

Likewise, the second term comes from $\tilde{F}_{p+1} = d A_p - \beta \wedge A_p+(\ldots)$. One possibility is that $\beta = q A_1$ for some one-form gauge field $A_1$, indicating that $A_p$ carries charge $q$ under $A_1$.\footnote{In this case, $\tilde{F}_{p+1}$ is only gauge-covariant, rather than gauge-invariant.} However, much like massless fermions, charged bosons neither affect black hole solutions nor mediate long-range forces at tree-level, because charge conservation disallows terms in the action containing only one charged field, and therefore all charged fields can be consistently truncated in a background preserving the gauge symmetry.\footnote{An exception is when the gauge symmetry is spontaneously broken close to the black hole, i.e., when the black hole solution passes onto another branch of the moduli space near the event horizon. Further discussion of this case is deferred to~\cite{CornersPaper}.}

Thus, for the time being we ignore all charged fields in the effective action. Once we have truncated the charged fields, $\beta = \beta_i(\phi) d \phi^i$ must be built from the scalar fields. Consistency of the Bianchi identity $d \tilde{F}_{p+1} = \beta \wedge \tilde{F}_{p+1} + \ldots$ implies that $d \beta - \beta \wedge \beta = 0$. Thus, $\beta$ is a flat one-form connection on the scalar manifold, implying that it is ``pure gauge,'' $\beta = -\Lambda^{-1} d \Lambda$ for some $\Lambda = \Lambda(\phi)$. Redefining $A_p \to \Lambda(\phi) A_p$ sets $\beta = 0$.

Therefore, after a field redefinition the neutral $p$-forms have modified field strengths,
\begin{align} \label{eqn:modField}
\tilde{F}_{p+1}^a &= d A_p^a + \gamma^a_{\; b} \wedge \tilde{A}_{p-1}^b + \ldots \,, &&\implies & d \tilde{F}_{p+1}^a &= \gamma^a_{\; b} \wedge \tilde{F}_p^b + \ldots \,,
\end{align}
where the additional terms involve wedge products of lower-rank forms and their exterior derivatives. Accounting for the fact that the modified Bianchi identities can contain terms with at most two derivatives at this order in the derivative expansion, we see that they take the general form
\begin{align} \label{eqn:modBianchi}
d \tilde{F}_{p+1}^a &= \sum_q C^a_{\;\; b c} \tilde{F}_{q+1}^b \wedge \tilde{F}_{p-q+1}^c \,,
\end{align}
for some constants $C^a_{\;\; b c}$.\footnote{A priori, $C^a_{\;\; b c}$ might depend on the scalar fields, but this is inconsistent with $d^2 \tilde{F}_{p+1}^a  = 0$.}

In terms of the modified field strengths, the two-derivative effective action for the neutral bosons is generally of the form:
\begin{align}
S &= \int d^d x \sqrt{-g} \biggl(\frac{1}{2 \kappa_d^2} R -\frac{1}{2} G_{i j}(\phi) \nabla\phi^i\cdot\nabla\phi^j -V(\phi) \biggr) 
- \sum_{p} \frac{1}{2} \int \f_{a b}(\phi) \tilde{F}_{p+1}^a \wedge \ast \tilde{F}_{p+1}^b \nonumber \\
 &\noeq + S_{\theta} + S_{\rm CS} \,, \label{eqn:generalaction}
\end{align}
where $V(\phi)$ is the scalar potential, $G_{i j}(\phi)$ is the metric on moduli space, and $\f_{ab}(\phi)$ is the gauge-kinetic matrix. We work in Einstein frame, so $\kappa_d^2$ is independent of the scalar fields, unlike $V$, $G$ and $\f$. The $\theta$ terms take the general form
\begin{equation}
 S_{\theta} = - \sum_{p} \frac{1}{8\pi^2} \int \theta_{a b}(\phi) \tilde{F}_{p+1}^a \wedge \tilde{F}_{d-p-1}^b \,, \label{eqn:thetaterm0}
\end{equation}
whereas the Chern-Simons interactions can be specified in a gauge-invariant manner as $S_{\rm CS} = - \int \cL_{\rm CS}$, where
\begin{equation}
d \cL_{\rm CS} = \sum_p \mu_{a b} \tilde{F}_{p+1}^a \wedge \tilde{F}_{d-p}^b + \sum_{p + q + r = d-2} \tilde{C}_{a b c} \tilde{F}_{p+1}^a \wedge \tilde{F}_{q+1}^b \wedge \tilde{F}_{r+1}^c 
\end{equation}
is a closed, gauge-invariant, formal $d+1$ form, involving at most three derivatives at this order. The two-derivative terms in $d \cL_{\rm CS}$ correspond to a one-derivative Chern-Simons terms. However, in combination with the usual Maxwell kinetic terms in~(\ref{eqn:generalaction}), these generate massive poles in the propagator. Since we integrated out all massive fields, we assume $\mu_{a b} = 0$.\footnote{Another possibility is a Chern-Simons gauge theory, where $\mu_{a b} \ne 0$ but $\f_{ab} = 0$. In this case, the equations of motion fix $\tilde{F}_{p+1}$ algebraically in terms of the other fields, fixing $A_p$ up to the addition of a flat gauge bundle. At the classical level, this gauge bundle has no effect on the black hole geometry. Understanding its quantum effects (if any) is an interesting question beyond the scope of this paper.}

The effect of the remaining cubic Chern-Simons interactions $\tilde{C}_{abc}$ on the equations of motion is easiest to see by defining the magnetic-dual field strengths
\begin{equation}
\tilde{F}_{q\; a}^{\text{(mag)}} \df 2 \pi \f_{a b} \ast \tilde{F}_{d-q}^b + \frac{\theta_{a b}}{2 \pi} \tilde{F}_{q}^b \,.
\end{equation}
In terms of the larger set of electric and magnetic field strengths the equations of motion and Bianchi identities combine into equations of the form~(\ref{eqn:modBianchi}). The couplings $C^a_{\;\;b c}$ and $\tilde{C}_{a b c}$ are therefore closely related and are intermixed by Hodge duality.

 Using Hodge duality, we can restrict to $1\le p \le (d-2)/2$. This eliminates most of the theta terms, except those
 for $p=(d-2)/2$ forms in even dimensions:
\begin{equation}
 S_{\theta} = - \frac{1}{8\pi^2} \int \theta_{a b}(\phi) \tilde{F}_{d/2}^a \wedge \tilde{F}_{d/2}^b \,. \label{eqn:thetaterm}
\end{equation}
An additional possibility in dimensions $d=4k+2$ is the presence of (anti)self-dual bosons, satisfying constraints of the form
\begin{equation} \label{eqn:selfdualityconstraint}
\ast \tilde{F}_{\frac{d}{2}}^a = \Lambda^a_{\; b}(\phi) \tilde{F}_{\frac{d}{2}}^b \,,
\end{equation}
for some $\Lambda^a_{\; b}(\phi)$. The resulting theory is non-Lagrangian: the constraints~\eqref{eqn:selfdualityconstraint} must be imposed by hand on top of the Euler-Lagrange equations for the ``pseudo-action''~\eqref{eqn:generalaction}.  

The arguments above are not meant to be rigorous, but provide a strong motivation for studying an effective action of the general form~(\ref{eqn:generalaction}). In the next section, I reduce the effective action to a simpler effective action that is equivalent for the purpose of characterizing spherically symmetric black holes and the long range forces between them. I then return to the issue of quantum effects, so far neglected.

\subsection{Static, spherically symmetric backgrounds} \label{eqn:sphericallysymmetric}

With the general form of the action in mind, let us specialize to static, spherically symmetric backgrounds, assumed henceforward.  Such backgrounds are sufficient to describe the long-range fields sourced by a particle at rest, and also to describe many (but not all) static black hole solutions.\footnote{As discussed in~\secref{subsec:extremalityDefn}, I will not consider multi-center solutions and other non-spherically symmetric black hole solutions.} 

Spherical symmetry severely restricts which background fields can be turned on, which allows us to truncate many fields and simplify the problem. In particular, the $\SO{d-1}$ rotational invariance around a particle or spherically symmetric black hole implies that $\tilde{F}_{p+1} = 0$ except in the Hodge-dual cases $p=1$ and $p=d-3$. Per~(\ref{eqn:modBianchi}), the $F_2$ Bianchi identity is unmodified, whereas a non-vanishing Chern-Simons contribution to $d \ast F_2$ could only involve $F_2 \wedge F_2$ (in $d=5$), but this too vanishes because $F_2 \propto d t \wedge d r$. We can therefore reduce the action~(\ref{eqn:generalaction}) to
\begin{equation} \label{eqn:sphericalBH}
S = \int d^d x \sqrt{-g} \biggl(\frac{1}{2 \kappa_d^2} R -\frac{1}{2} G_{i j}(\phi) \nabla\phi^i\cdot\nabla\phi^j -V(\phi) \biggr)
- \frac{1}{2} \int \f_{a b}(\phi)  F_2^a \wedge \ast F_2^b + S_{\theta} 
\end{equation}
without affecting the subsequent calculations in this paper.

Consistent with a Minkowski vacuum, suppose that $V(\phi) \to 0$ asymptotically far from the black hole, with $V(\phi) \ge 0$ nearby in scalar field space.
Because the scalar potential is lower-order in the derivative expansion than other terms in the action, for parametrically large black holes it is parametrically important. The associated force pushes solutions down to the moduli space $V(\phi) = 0$ in the large black hole limit, hence we can ignore all scalar fields that are not moduli in this limit, restricting to the submanifold $V(\phi) = 0$ of scalar field space.\footnote{Massless scalars that are not moduli contribute to long-range forces, but since parametrically large black holes are confined (very close) to the moduli space, they are not charged under these scalars.}

The theta-term $S_{\theta} = - \frac{1}{8\pi^2} \int \theta_{a b}(\phi) F_2^a \wedge F_2^b$ is present only in 4d, in which case the black hole can carry magnetic charge $F_2 \propto \Omega_2$ as well as electric charge $F_2 \propto dt \wedge dr$, where $\Omega_2$ is the volume form of the transverse $S^2$. If the magnetic charge vanishes, then $F_2 \propto dt \wedge dr$, hence $d \phi \wedge F_2 = 0$ because $d \phi \propto d r$ by spherical symmetry. Thus, for purely electrically charged particles and black holes, the theta term can be ignored. I focus on this case in the main text for simplicity, deferring a complete treatment of magnetic charge to appendix~\ref{app:democratic}.

With this caveat, we see that the following action is sufficient for analyzing parametrically large spherically symmetric black holes and long-range forces between them
\begin{equation} \label{eqn:gen-action-BH}
S = \int d^d x \sqrt{-g} \biggl(\frac{1}{2 \kappa_d^2} R -\frac{1}{2} G_{i j}(\phi) \nabla\phi^i\cdot\nabla\phi^j - \frac{1}{2} \f_{a b}(\phi) F_2^a \cdot F_2^b \biggr) \,,
\end{equation}
where we use
\begin{equation}
\int \omega_p \wedge \ast \chi_p = \int d^d x \sqrt{-g}\, \omega_p \cdot \chi_p \qquad \text{for} \qquad \omega_p \cdot \chi_p \df \frac{1}{p!} \chi_{M_1 \ldots M_p} \omega^{M_1 \ldots M_p} \,. \label{eqn:dotdefn}
\end{equation}

\subsubsection{Black branes} \label{subsec:branesetup}

Similar considerations apply to black branes and the long range forces between them. I focus on $(p-1)$-branes with $1 < p \le d-3$ that are rotationally invariant in the directions transverse to their worldvolume and invariant under spacetime translations and spatial rotations along their worldvolume. The complete symmetry group is then
\begin{equation} \label{eqn:non-boost-inv-symms}
\SO{d-p}\times \mathbb{R} \times \ISO{p-1} \,,
\end{equation}
where the Euclidean group $\ISO{p-1}$ acts on the worldvolume spatial directions and $\mathbb{R}$ translates in time.

We could also demand boost-invariance (hence, Poincar\'e invariance) along the worldvolume, in which case the symmetry group would be
\begin{equation}
\SO{d-p}\times \ISO{p-1,1} \,. \label{eqn:boost-inv-symms}
\end{equation}
Such a brane cannot truly be ``black'' because 
boost invariance requires $g_{\mu \nu} \to 0$ for all worldvolume directions $\mu, \nu$ whenever $g_{t t} \to 0$, making a smooth horizon impossible. However, quasiextremal solutions can be boost-invariant, and in fact they always are with the assumptions used in this paper.

In the boost-invariant case, the symmetries (\ref{eqn:boost-inv-symms}) impose $\tilde{F}_{q+1} = 0$ except for $q=p$, $q=p-1$, and their Hodge duals $q=d-p-2$ and $q = d-p-1$.
 In particular,
\begin{equation}
\tilde{F}_p = f(r) dt\wedge d y^1 \wedge \ldots \wedge dy^{p-1} \,,
\end{equation}
but the Bianchi identity~\eqref{eqn:modBianchi} imposes $f'(r) = 0$ because the symmetries do not allow $\tilde{F}_{q+1} \wedge \tilde{F}_{r+1}$ with $q,r\ge 1$ to have a component along $dt\wedge d y^1 \wedge \ldots \wedge dy^{p-1} \wedge d r$. Thus, $\tilde{F}_p$ retains a constant background value far from the brane. This is inconsistent with asymptotic flatness, so we require $\tilde{F}_p = 0$.

The action for the remaining fields is
\begin{equation} \label{eqn:pformaction0}
S = \int d^d x \sqrt{-g} \biggl(\frac{1}{2 \kappa_d^2} R -\frac{1}{2} G_{i j}(\phi) \nabla\phi^i\cdot\nabla\phi^j -V(\phi) \biggr)
- \frac{1}{2} \int \f_{a b}(\phi)  F_{p+1}^a \wedge \ast F_{p+1}^b + S_{\theta} \,,
\end{equation}
analogous to~\eqref{eqn:sphericalBH}. To verify that this is a consistent truncation, note that the symmetries require $\tilde{F}_{p+1}^a \propto dt\wedge d y^1 \wedge \ldots \wedge dy^{p-1} \wedge d r$ except when $d=2p+2$, where a component along $S^{d-p-1}$ is also possible. Thus, $\tilde{F}_{p+1}^a \wedge \tilde{F}_{(d-p-1)\; b}^{(mag)}$ and (for $d=2p+2$) $\tilde{F}_{d/2}^a \wedge \tilde{F}_{d/2}^b$ and $\tilde{F}_{\frac{d}{2}\; a}^{(mag)} \wedge \tilde{F}_{\frac{d}{2}\; b}^{(mag)}$ are the only non-vanishing wedge products of $q$-form gauge fields in this background. As top forms, these cannot appear as source terms in the $\tilde{F}_{q+1}^a$ Bianchi identities and equations of motion for $0 < q < d-2$, allowing these fields to be consistently truncated for $q \ne p$.

With the smaller symmetry group~\eqref{eqn:non-boost-inv-symms}, the black brane can carry a one-form charge density sourcing $F_2 \propto d t \wedge d r$ and a $(p-1)$-form charge density sourcing $\tilde{F}_p \propto dy^1 \wedge \ldots \wedge dy^{p-1} \wedge d r$ in addition to its $p$-form charge. I will focus on the case where these charge densities vanish, in which case $F_2 = \tilde{F}_p = 0$, and we can again consistently truncate to the action~\eqref{eqn:pformaction0}. Note that in this case the larger symmetry group~\eqref{eqn:boost-inv-symms} is broken by the metric only, and not by the other fields.

The theta term~\eqref{eqn:thetaterm} is only present in $d=2p+2$ dimensions. As above, it has no effect on black $(p-1)$-branes with purely electric charge. For simplicity, I focus on this case in the main text, leaving further discussion of magnetic charges, theta angles, and self-dual gauge fields to appendix~\ref{app:democratic}.

With these caveats, the action simplifies to (see~\eqref{eqn:dotdefn})
\begin{equation} \label{eqn:pformaction}
S = \int d^d x \sqrt{-g} \biggl(\frac{1}{2 \kappa_d^2} R -\frac{1}{2} G_{i j}(\phi) \nabla\phi^i\cdot\nabla\phi^j - \frac{1}{2} \f_{a b}(\phi) F_{p+1}^a \cdot F_{p+1}^b \biggr) \,,
\end{equation}
sufficient for analyzing black $(p-1)$-branes with the indicated symmetries and the long-range forces between them. 

\subsection{Singularities in the moduli space} \label{subsec:MSsingularities}

Until now we have neglected all quantum effects. For $d>4$ all interactions between massless fields are irrelevant\footnote{Although a cubic scalar interaction is relevant (marginal) in $d=5$ ($d=6$), the presence of such an interaction between massless scalars implies that $V(\phi)$ decreases after an infinitesimal displacement along some direction in field space, and we are not in a stable vacuum.} and the theory is infrared free. Because of this, loop contributions will generally be subleading in comparison with the tree-level interactions, and so the latter will completely determine the leading long-range forces and the leading behavior of parametrically large black holes.

However, an important subtlety occurs when a massive particle becomes massless along some locus in the moduli space.\footnote{I thank M.~Alim, M.~Reece and T.~Rudelius for extensive discussions on related points. Some associated results are reported in~\cite{conifolds}.} Moving infinitesimally off this locus and integrating out the now massive particle, loop effects will generate new terms in the effective action for the modulus along which we displaced and other fields coupled to it. These new couplings are generally \emph{non-analytic} functions of the modulus, and so cannot be absorbed into tree-level counterterms. By comparison, while the same motion will generally change the masses of various already-massive fields, the loop effects associated with these changes are analytic at the chosen point in moduli space, and can be absorbed by counterterms.

Thus, the functions $\f_{a b}(\phi)$ and $G_{i j}(\phi)$ should be analytic functions of the moduli away from certain singular points in the moduli space where additional particles become massless. Moreover, the non-analytic behavior of $\f_{a b}(\phi)$ and $G_{i j}(\phi)$ at the singular points should correspond to loops of the particles becoming massless there.\footnote{I assume that $V(\phi) = 0$ is maintained at loop level, e.g., due to supersymmetry.}

\begin{figure}
\begin{center}
\begin{subfigure}{0.475\textwidth}
\center
\includegraphics{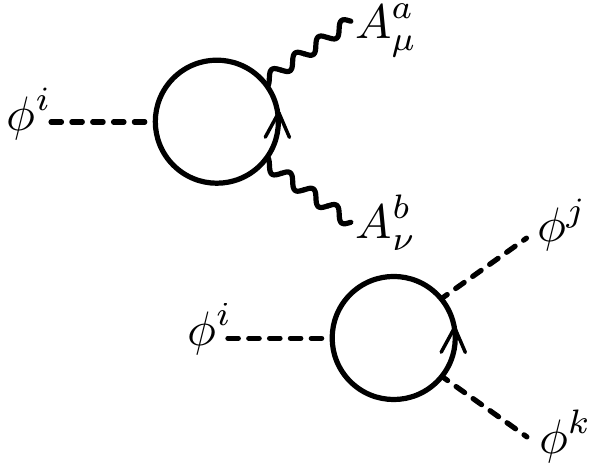}
\caption{One-loop corrections to $\f_{a b}(\phi)$, $G_{j k}(\phi)$ \\ \ } \label{sfig:OneLoop}
\end{subfigure}
\hfill
\begin{subfigure}{0.475\textwidth}
\center
\includegraphics{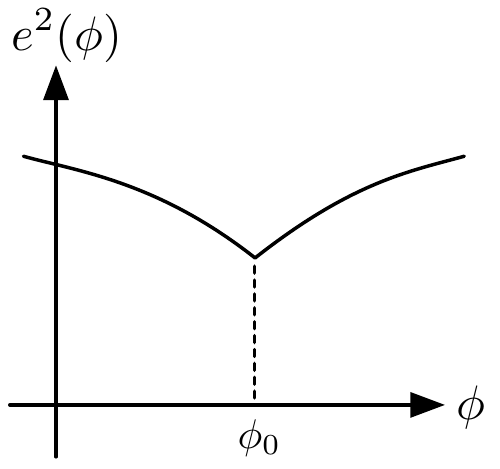}
\caption{Singular point in a 5d theory due to massless charged particles} \label{sfig:SingularPoint}
\end{subfigure}
\caption{\subref{sfig:OneLoop}~Leading contributions to non-analytic couplings near a singular point in the moduli space of a 5d theory, in the case where the particle becoming massless is a fermion (similar diagrams apply to bosons). \subref{sfig:SingularPoint}~Behavior of a 5d gauge coupling $e^2(\phi) = 1/\f(\phi)$ near a singular point in the moduli space $\phi = \phi_0$ where massless charged particles appear.\label{fig:singularpoint}}
\end{center}
\end{figure}

As an example, in 5d a particle of mass $M(\phi) = \mu_i \phi^i + O(\phi^2)$ and charge $Q_a$ contributes non-analytic couplings of the form $\Delta f_{a b}(\phi) \propto Q_a Q_b |\mu_i \phi^i |$ and $\Delta G_{i j}(\phi) \propto \mu_i \mu_j |\mu_k \phi^k |$ near the point $\phi^i = 0$, as illustrated in Figure~\ref{fig:singularpoint}. Specific $d=5$ quantum gravities of this kind are analyzed in detail in~\cite{conifolds}. In larger dimensions, the leading non-analytic behavior is higher-order in the moduli displacement.

The situation in 4d is subtly different. Because gauge interactions are marginal, the presence of massless charged particles (or massless non-Abelian gauge fields) has a larger effect. If there are massless charged particles everywhere in moduli space then generically the gauge couplings will either run to zero in the infrared or become non-perturbatively large, depending on the signs of their beta functions. Either way, quantum effects play an essential role in the deep infrared. Analyzing such a situation is beyond the scope of this paper.\footnote{The non-perturbatively large couplings we are concerned with here do not include those associated with confining gauge theories, since the confined gauge symmetry plays no role in the deep infrared.}

\begin{figure}
\begin{center}
\begin{subfigure}{0.475\textwidth}
\center
\includegraphics{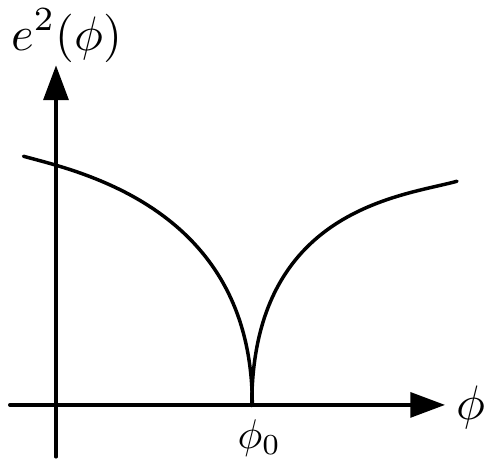}
\caption{Singular point in a 4d theory due to massless charged particles} \label{sfig:SingularPoint4D}
\end{subfigure}
\hfill
\begin{subfigure}{0.475\textwidth}
\center
\includegraphics{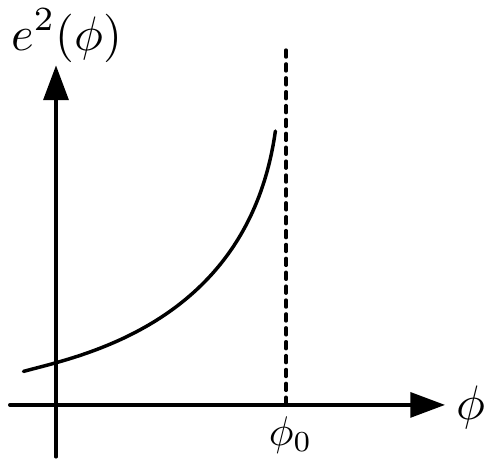}
\caption{A strongly-coupled singular point \\ \ } \label{sfig:SingularPointCFT}
\end{subfigure}
\caption{\subref{sfig:SingularPoint4D}~In 4d theories massless charged particles lead to logarithmic infrared divergences, driving the gauge coupling to zero in the deep infrared. \subref{sfig:SingularPointCFT}~A singular point at which perturbation theory breaks down, signaled by diverging or large dimensionless couplings nearby.\label{fig:moresingularpoints}}
\end{center}
\end{figure}

Suppose instead that massless charged particles only appear at loci in the moduli space of codimension one or higher. In this case, the interactions are irrelevant away from these singular loci, and we can continue to use the effective action~\eqref{eqn:pformaction} everywhere else in the moduli space. The main difference versus higher-dimensional theories is that in 4d the singularities in~$\f_{ab}(\phi)$ and $G_{i j}(\phi)$ are more dramatic, see Figure~\ref{sfig:SingularPoint4D}.

A further possibility in both 4d theories and higher-dimensional theories is that perturbation theory breaks at certain points in the moduli space, see Figure~\ref{sfig:SingularPointCFT}. This can happen due to the appearance of a conformal field theory (CFT), or when a monopole or dyon becomes massless (as in Seiberg-Witten theory~\cite{Seiberg:1994rs}).

\subsubsection{Effect on charged black hole solutions}

Because the moduli travel within the moduli space as we approach the event horizon, sometimes by a substantial distance, charged black hole solutions can be sensitive to distant features in the moduli space. 
To understand the effects of moduli space singularities on black hole solutions, consider for example a 4d theory with a $\U{1}$ gauge group and a one-dimensional moduli space in which a charged particle becomes massless at a singular point $\phi = \phi_0$, see Figure~\ref{sfig:SingularPoint4D}. Suppose we are interested in black hole solutions in a vacuum $\phi_{\infty} = \phi(r = \infty)$, where $\phi_{\infty}$ is close to (but not at) the singular point $\phi_0$. Because the gauge coupling is minimized at $\phi = \phi_0$, the modulus is drawn towards this point near the event horizon, reducing the electrostatic energy $|\vec{E}|^2 \propto e^2 Q^2$.\footnote{This is qualitative explanation, but the same result can be shown explicitly using~\eqref{eqn:smoothhorizon} and \eqref{eqn:phizEOM}.} In the quasiextremal limit, the value at the horizon $\phi_h$ goes to $\phi_0$ due to the attractor mechanism~\cite{Ferrara:1995ih,Cvetic:1995bj,Strominger:1996kf,Ferrara:1996dd,Ferrara:1996um}.

\begin{figure}
\begin{center}
\begin{subfigure}{0.475\textwidth}
\center
\includegraphics{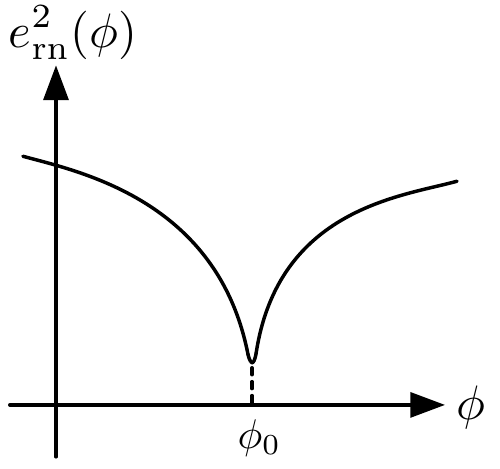}
\caption{Renormalized singular point} \label{sfig:SingularPointRenormalized}
\end{subfigure}
\hfill
\begin{subfigure}{0.475\textwidth}
\center
\includegraphics{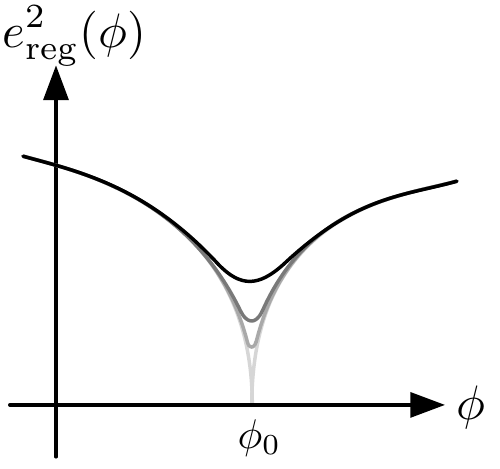}
\caption{Limit of analytic functions} \label{sfig:SingularPointLimit}
\end{subfigure}
\caption{\subref{sfig:SingularPointRenormalized}~The gauge coupling renormalized at the scale of the horizon radius avoids the singular behavior at $\phi = \phi_0$. \subref{sfig:SingularPointLimit}~The non-analytic behavior near a singular point can be described as the limit of a sequence of analytic functions, appropriate to the large black hole limit.}
\end{center}
\end{figure}

Naively, the gauge coupling is exactly zero at the singular point $\phi = \phi_0$, due to the screening effect of the massless charged particle. However, for a finite size black hole the size of the near-horizon region is likewise finite, and consequently the gauge coupling is not completely screened. Defining an appropriate renormalized coupling to account for the finite size of the black hole, the singularity in moduli space is smoothed, see Figure~\ref{sfig:SingularPointRenormalized}. This is an example of the well-known fact that phase transitions do not occur in a finite volume, thus finite size black holes do not probe actual singularities in the moduli space, but rather very sharp, analytic features that approximate them.

Honestly calculating these finite-size effects requires a much more careful treatment of quantum effects beyond the scope of this paper. However, we can crudely model them by postulating some \emph{analytic} functions $\f_{a b}^{\rm reg}(\phi)$ and $G_{i j}^{\rm reg}(\phi)$ (roughly speaking the renormalized couplings) that closely approximate $\f_{a b}(\phi)$ and $G_{i j}(\phi)$ away from the singular points. As the black hole size is increased, the appropriate $\f_{a b}^{\rm reg}(\phi)$ and $G_{i j}^{\rm reg}(\phi)$ should be made sharper near the singularities. Thus, in the large black hole limit, we take a limit of analytic functions $\f_{a b}^{\rm reg}(\phi) \to \f_{a b}(\phi)$ and $G_{i j}^{\rm reg}(\phi) \to G_{i j}(\phi)$, recovering the non-analytic behavior as the black hole size goes to infinity, as illustrated in Figure~\ref{sfig:SingularPointLimit}.

With this in mind, to construct black hole solutions we begin with the simplifying assumption that $\f_{a b}(\phi)$ and $G_{i j}(\phi)$ are analytic. Singularities in the moduli space can then be described by taking a limit of these solutions as $\f_{a b}(\phi)$ and $G_{i j}(\phi)$ develop localized sharp features. Typically the result of taking such a limit will not depend on the choice of a sequence of analytic functions $\f_{a b}^{\rm reg}(\phi)$ and $G_{i j}^{\rm reg}(\phi)$ approaching the desired singular $\f_{a b}(\phi)$ and $G_{i j}(\phi)$, and so we can describe parametrically large black hole solutions parametrically well without needing to analyze the aforementioned finite-size effects in detail. 

\subsubsection{Strongly coupled black holes?}

Above we focused on singularities with a perturbative description. One might worry that a strongly-coupled singularity (e.g., Figure~\ref{sfig:SingularPointCFT}) might lead to a loss of calculability, i.e., knowledge of the infrared couplings $\f_{a b}(\phi)$ and $G_{i j}(\phi)$ might be insufficient to determine the large black hole solutions. However, for the same reason that they are attracted to regions with small gauge couplings, ordinary charged black hole solutions are generally repelled from regions with large gauge couplings, and so are insensitive to the physics of the strongly-coupled singular points.

Despite this, it is possible in principle that a separate class of ``strongly coupled'' black hole solutions exist, with strongly-interacting degrees of freedom appearing a finite distance outside the event horizon. Such solutions---if they exist\footnote{One obstacle to finding such solutions is that the large gauge coupling leads to a large electrostatic contribution to the stress tensor, potentially leading to a naked singularity, as it does for a superextremal RN ``black hole.'' This requires more study,  left to future work.\label{fn:stronglycoupledBHs}}---cannot be analyzed by an effective field theory of the form described above, and are outside the scope of this paper.

\section{Long-range forces} \label{sec:forces}

At very low energies (i.e., at length scales large compared to the event horizon), a black hole looks just like a massive particle. Effective field theory principles then imply that the low-energy dynamics of a black hole can be described by some point particle effective action. In particular, this action accurately describes the force on a black hole in the slowly varying background of another distant black hole. Thus, to determine the long range force between a pair of charged black holes, we consider the long range force between a pair of massive, charged particles. The latter turns out to depend only on mass $M$, charge $Q$, and scalar charge $\frac{dM}{d\phi}$, and therefore the long range force between charged black holes is determined by the same three quantities in the same way.

Similar reasoning applies to black branes with membranes in place of particles, with a few complications to be discussed below. Thus, to determine the long range force between black branes, we study the analogous forces between dynamical membranes. 
Both particle and membrane cases are treated in turn in the following sections.

\subsection{Particles}
\label{sec:particleselfforce}

Let us begin with particles.
The probe action for a massive particle in slowly varying background fields is
\begin{equation} \label{eqn:worldlineS}
  S_{\mathrm{pp}} = - \int M (\phi)\, d s - Q \int A \; .
\end{equation}
This is a slight generalization of the usual action for a massive particle (see, e.g.,~\cite{Landau}, chapter 3), where the mass is allowed to depend on a background modulus $\phi$. The form of the action is completely fixed by the symmetries up to terms involving derivatives of the background fields. In particular, $Q$ cannot depend on $\phi$, as this would violate gauge invariance. We omit any terms involving derivatives of the background fields, as these do not contribute to the long range forces.

Choosing a worldline parameterization $\lambda$, we define the projection operator
\begin{equation}
  P^{\rho}_{\;\nu} \df \delta^{\rho}_{\;\nu} - \frac{1}{\dot{x}^2} 
  \dot{x}^{\rho}  \dot{x}_\nu \;,\qquad \dot{x}^\mu \df \frac{d x^\mu}{d\lambda} \, ,
\end{equation}
which projects onto spatial components in the particle rest frame.
In terms of this, the equation of motion is
\begin{equation}
  P^{\mu}_{\;\nu}  \left[ \frac{M (\phi)}{\sqrt{- \dot{x}^2}}  (\ddot{x}^{\nu} +
  \Gamma^{\nu}_{\;\rho \sigma}  \dot{x}^{\rho}  \dot{x}^{\sigma}) + M'
  (\phi)  \sqrt{- \dot{x}^2} \nabla^{\nu} \phi + Q F^{\nu}_{\;\;\rho}  \dot{x}^{\rho} \right] = 0 \; .
  \label{eqn:partEOM}
\end{equation}
Under worldline reparameterizations, $\ddot{x}^\nu$ picks up a piece proportional to $\dot{x}^\nu$; this is annihilated by the projection operator, ensuring the covariance of the equation.
Fixing the parameterization $\lambda = \tau$ with $\tau$ the worldline proper time (so that $\dot{x}^2 = -1$), the equation of motion simplifies to
\begin{equation}
  M (\phi) (\ddot{x}^{\mu} + \Gamma^{\mu}_{\;\nu \rho}  \dot{x}^{\nu} 
  \dot{x}^{\rho}) + M' (\phi) P^{\mu}_{\;\nu} \nabla^{\nu} \phi +  Q F^{\mu}_{\;\;\nu}  \dot{x}^{\nu} = 0 \;, \qquad
  \dot{x}^2 = - 1 \;, \label{eqn:taueom}
\end{equation}
where now $P^{\mu}_{\;\nu} = \delta^{\mu}_{\;\nu} + \dot{x}^{\mu} 
\dot{x}_{\nu}$.

Given a Killing vector $K_\mu$ (satisfying $\nabla_{(\mu} K_{\nu)} = 0$) such that $K^{\mu} \nabla_{\mu} \phi = 0$ and $K^{\mu} F_{\mu \nu} = -
\nabla_{\nu} U$ for some potential $U$, the equation of motion~(\ref{eqn:taueom}) implies a conservation law
\begin{equation}
  \frac{d}{d \tau}  \bigl[-M (\phi) K_{\nu}  \dot{x}^{\nu} + Q U\bigr] = 0 \; .
  \label{eqn:generalCons}
\end{equation}
In particular, in a static background $d s^2 = g_{t t}(x) d t^2 + g_{i j} (x) d x^i d x^j$, $\phi = \phi(x)$ and $A = \Phi (x) d t$, we find the conserved energy
\begin{equation}
  E = M (\phi)  \sqrt{- g_{t  t}  (1 + g_{i j}  \dot{x}^i  \dot{x}^j)} + Q \Phi \,.\label{eqn:conservedE}
\end{equation}
Note that the motion is integrable in a spherically symmetric background, $d s^2 = g_{t t} (r) d t^2 + g_{r r} (r) d r^2 + R (r)^2 d \Omega_{D - 2}^2$, $\phi = \phi(r)$, and $\Phi = \Phi(r)$. In particular,
\begin{equation}
 g_{r r}  \dot{r}^2 + 1 + \frac{(E - Q \Phi)^2}{g_{tt} M^2 (\phi)} +
  \frac{L^2}{R(r)^2 M (\phi)^2} = 0  \;,
\end{equation}
where $E > Q \Phi$ and $L^2 \ge 0$ is the conserved total angular momentum.

Taking the non-relativistic ($|\dot{\boldsymbol{x}}| \ll 1$) and weak-field ($g_{\mu \nu} = \eta_{\mu \nu} + h_{\mu \nu}$ for $| h_{\mu
\nu} | \ll 1$) limit, we obtain:
\begin{equation}
  E_{\mathrm{nr}} \simeq \frac{1}{2} M (\phi) \dot{\boldsymbol{x}}^2 + (1+\psi) M (\phi) + Q \Phi \,,
  \label{eq:Enr}
\end{equation}
where $\psi = - h_{t t}/2$ is the Newtonian potential. Thus, we identify $V_{\mathrm{nr}} = (1+\psi) M (\phi) + Q \Phi$ as the non-relativistic potential energy.
This specifies the force on a massive charged particle probing a fixed, slowly varying background. 

\subsubsection{Linearized backreaction}
To calculate the long-range force between two massive charged particles, we compute the linearized background fields sourced by the action~(\ref{eqn:worldlineS}).
First, we introduce an action for the background fields:
\begin{equation}
  S_{\mathrm{fields}} = \int d^d x \sqrt{- g}  \left[ \frac{1}{2 \kappa_d^2}
  R- \frac{1}{2} G_{\phi \phi}  (\nabla \phi)^2 -
  \frac{1}{4 e^2} F_{\mu \nu} F^{\mu \nu} \right] \; .
\end{equation}
In general, $G_{\phi \phi}$ and $e^2$ will be functions of the modulus, $\phi$, but as we will be interested in the linearized equations of motion, we ignore this dependence.

From $S = S_{\mathrm{fields}} + S_{\mathrm{pp}}$, we obtain the field equations
\begin{align}
  \frac{1}{2 \kappa_d^2} \biggl(R^{\mu \nu}-\frac{1}{2} g^{\mu \nu} R \biggr)  + (\ldots) &=  \int \frac{\delta^{(d)} (x -
  x(\lambda))}{\sqrt{- g}}  \frac{\dot{x}^{\mu} \dot{x}^{\nu}}{2
  \sqrt{-\dot{x}^2}} M (\phi) d \lambda \,, \nonumber \\
  G_{\phi \phi} \nabla^2 \phi &= \int \frac{\delta^{(d)} (x -
  x(\lambda))}{\sqrt{- g}}  \sqrt{- \dot{x}^2} M' (\phi) d \lambda
  \,, \nonumber \\
  \frac{1}{e^2} \nabla_{\mu} F^{\mu \nu} &= Q \int \frac{\delta^{(d)}
  (x - x (\lambda))}{\sqrt{- g}} \; \dot{x}^{\nu} d \lambda \, , \label{eqn:backEOM}
\end{align}
where the omitted terms in the Einstein equations depend quadratically on the background fields $\nabla \phi$ and $F_{\mu \nu}$.
In particular, for a particle stationary at the origin $x^i = 0$, (\ref{eqn:backEOM}) becomes
\begin{align}
  R^{\mu \nu}-\frac{1}{2} g^{\mu \nu} R + (\ldots) &= \kappa_d^2 M (\phi)  \frac{\delta^{(d - 1)} (x^i)}{\sqrt{-g}} \delta^{\mu t} \delta^{\nu t} \,, \nonumber \\
  G_{\phi \phi} \nabla^2 \phi &= \frac{\delta^{(d - 1)}
  (x^i)}{\sqrt{- g}} M' (\phi) \,,  &
  \frac{1}{e^2} \nabla_{\mu} F^{\mu \nu} &= Q \frac{\delta^{(d - 1)}
  (x^i)}{\sqrt{- g}} \; \delta^{\nu t} \, .  
\end{align}
Linearizing around a background $g_{\mu \nu} = \eta_{\mu \nu}$ and $\phi =
\phi_0$, we obtain
\begin{align}
  - \frac{1}{2} \partial^2 \bar{h}^{\mu \nu} &= \kappa_D^2 M (\phi_0)
  \delta^{(d - 1)} (x^i) \delta^{\mu t} \delta^{\nu t} \,, \nonumber \\
  G_{\phi \phi} \partial^2 \phi &= M' (\phi_0) \delta^{(d - 1)}
  (x^i) \,, &
  \frac{1}{e^2} \partial_{\mu} F^{\mu \nu} &= Q \delta^{(d - 1)} (x^i) \,
  \delta^{\nu t} \, ,
\end{align}
where $g_{\mu \nu} = \eta_{\mu \nu} + h_{\mu \nu}$, $\bar{h}_{\mu \nu} \df h_{\mu \nu} - \frac{1}{2} \eta_{\mu \nu} h$, and we work in Lorenz gauge $\partial_{\mu}  \bar{h}^{\mu \nu} = 0$.\footnote{See, e.g.,~\cite{Schutz:1985jx} section 8.3 for an overview of linearized gravity.}

In $(d - 1)$ spatial dimensions,
\begin{align} \label{eqn:CoulombDdims}
  \partial_i \partial^i  \frac{1}{r^{d - 3}} &= - (d - 3) V_{d - 2} \delta^{(d - 1)} (x^i) \,, & V_{d - 2} &= \frac{2 \pi^{\frac{d - 1}{2}}}{\Gamma \bigl( \frac{d - 1}{2} \bigr)} \,,
\end{align}
where $V_{d-2}$ is the volume of $S^{d - 2}$ and we use Gauss's law to fix the normalization.
Thus, we obtain the static, spherically symmetric solution:
\begin{align}
  \bar{h}_{t t} &= \frac{2 \kappa_d^2 M (\phi_0) }{(d - 3) V_{d - 2}} \frac{1}{r^{d - 3}} , & \phi &= \phi_0 - \frac{G_{\phi \phi}^{- 1} M' (\phi_0)}{(d - 3) V_{d - 2}}  \frac{1}{r^{d - 3}} , &
  \Phi &= \frac{e^2 Q}{(d - 3) V_{d - 2}}  \frac{1}{r^{d - 3}} ,
\end{align}
at the linearized level. The Newtonian potential $\psi$ is
\begin{equation}
  \psi = - \frac{d-3}{2 (d - 2)} \bar{h}_{t t} = - \frac{\kappa_d^2 M (\phi_0) }{(d - 2) V_{d - 2}}  \frac{1}{r^{d - 3}} \; .
\end{equation}
Applying~(\ref{eq:Enr}), we find the potential between two widely separated, non-relativistic particles:
\begin{equation}
  V_{\mathrm{nr}} = - \frac{\kappa_d^2 }{(d - 2) V_{d - 2}}  \frac{M_1 M_2}{r^{d - 3}} - \frac{G_{\phi \phi}^{-1}}{(d - 3) V_{d - 2}} \frac{\partial_{\phi} M_1 \partial_{\phi} M_2}{r^{d - 3}}
  + \frac{e^2}{(d-3) V_{d - 2}}  \frac{Q_1 Q_2}{r^{d - 3}} \; .
\end{equation}
The magnitude of the long-range central force is therefore
\begin{equation}
  F_{12} = - \frac{(d-3) \kappa_d^2 }{(d - 2) V_{d - 2}}  \frac{M_1 M_2}{r^{d - 2}} - \frac{G_{\phi \phi}^{-1}}{V_{d - 2}} \frac{\partial_{\phi} M_1 \partial_{\phi} M_2}{r^{d - 2}}
  + \frac{e^2}{V_{d - 2}}  \frac{Q_1 Q_2}{r^{d - 2}} \; ,
\end{equation}
with $F_{12} > 0$ for a repulsive force.

Note that the scalar charge $\partial_{\phi} M$ is more correctly written $M_{d} \partial_{\phi} \frac{M}{M_d}$, where $M_d = \kappa_d^{-2/(d-2)}$ is the $d$-dimensional Planck mass. In particular, this can differ from $M_0 \partial_{\phi} \frac{M}{M_0}$ for some other mass scale $M_0$ (such as the $D=d+1$ dimensional Planck scale in Kaluza-Klein theory) because $M_0 / M_d$ may depend on $\phi$. In general, moduli derivatives of dimensionful quantities are ambiguous until we specify which scale is held fixed.

This calculation is easily generalized to the case with multiple gauge bosons and moduli. Fixing the background field action \eqref{eqn:gen-action-BH}
\begin{equation}
  S_{\mathrm{fields}} = \int d^d x \sqrt{- g}  \left[ \frac{1}{2 \kappa_d^2} R - \frac{1}{2} G_{i j} (\phi) \nabla \phi^i \cdot \nabla \phi^j - \frac{1}{4} \f_{a b}(\phi) F^a_{\mu \nu} F^{b \mu \nu} \right] \;,
\end{equation}
the long range force between two particles comes out to
\begin{align} \label{eqn:particleForce}
F_{12} &= \frac{\mathcal{F}_{12}}{V_{d-2}\, r^{d-2}} , & \mathcal{F}_{12} = \f^{a b} Q_{1a} Q_{2b} - G^{i j} \mu_{1 i} \mu_{2 j}- \frac{d-3}{d - 2} \kappa_d^2 M_1 M_2 \,,
\end{align}
where $\f^{a b}$ and $G^{i j}$ are the inverse gauge kinetic matrix and scalar metric, respectively, 
and $\mu_i \df \frac{\partial M}{\partial \phi^i}$ is the scalar charge.

\subsection{Dirac branes}
\label{sec:braneselfforce}

Now consider membranes.\footnote{I thank M.~Reece and T.~Rudelius for helpful discussions and initial collaboration on this topic.} In the special case that a $(p-1)$-brane is characterized solely by its tension $\cT$ and $p$-form charge $Q$, it has an essentially unique action
\begin{equation} \label{eqn:Diracaction}
S_p = - \int d^p \xi  \sqrt{-\tilde{g}}\, \cT(\phi) - Q \int A_p
\end{equation}
at leading order in slowly varying background fields, where $\tilde{g}$ is the pullback of the spacetime metric $g$ to the brane worldvolume. 
This is the charged Dirac membrane action, matching the familiar form of BPS brane actions in string theory and 
generalizing the point-particle action~(\ref{eqn:worldlineS}). 

The Dirac action is boost-invariant along the brane worldvolume, hence it can describe black branes with the full symmetry group \eqref{eqn:boost-inv-symms}. However, non-extremal black branes (whose mass density and tension generally differ) are not boost-invariant along their worldvolumes, and require a more general approach.\footnote{As non-extremal black branes suffer from the Gregory-Laflamme instability~\cite{Gregory:1993vy,Gregory:1994bj,Gregory:1994tw}, they may not live long enough to measure the force between them. However, nearly extremal black branes should be long lived, and for this reason I will not explore this issue further.}
For simplicity, I focus on boost-invariant Dirac branes in this section, returning to the general case (with reduced symmetry~\eqref{eqn:non-boost-inv-symms}) later on.

Consider a background of the form $d s^2 = g_{\mu \nu} d x^{\mu} d x^{\nu} + g_{m n} d x^m d x^n$ and $A_p = \Phi d x^0 \wedge \ldots \wedge d x^{p-1}$, where $\mu, \nu = 0, \ldots, p-1$, and $m, n = p, \ldots d-1$.
Fixing static gauge, $\xi^\mu = x^\mu$, and expanding in small fluctuations about $x^m(\xi) = 0$, we find
\begin{equation}
S_p \approx \int d^p \xi \sqrt{-\det g_{\mu \nu}}\, \cT(\phi) \biggl[-\frac{1}{2} g_{m n} g^{\mu \nu} \partial_\mu x^m \partial_\nu x^n - 1 \biggr] - Q \int d^p \xi\, \Phi \,.
\end{equation}
In the weak field limit, where $g_{\mu \nu} = \eta_{\mu \nu} + h_{\mu \nu}$ and $g_{m n} = \delta_{m n} + h_{m n}$, this reduces to
\begin{equation}
S_p \approx \int d^p \xi \, \biggl[\frac{1}{2} \cT(\phi) (\dot{x}^m)^2 -\frac{1}{2} \cT(\phi) (\partial_i x^m)^2 - \cT(\phi) (1+ \Psi_p) - Q \Phi \biggr] \,,
\end{equation}
where the $p$-brane analog of the Newtonian potential $\Psi_p$ is:
\begin{equation}
\Psi_p = \frac{1}{2} \eta^{\mu \nu} h_{\mu \nu} = \frac{1}{2} (-h_{t t} + \delta^{i j} h_{i j}) \,.
\end{equation}
Note that $\cT(\phi)$ simultaneously plays two distinct roles in this action. It is the energy density of the brane as well as its tension. The gravitational coupling $\cT(\phi) \Psi_p$ accounts for both: as we will show later, $-h_{t t}/2$ couples to the mass density, whereas $\delta^{i j} h_{i j}/2$ couples to the tension.

We read off the potential energy density:
\begin{equation} \label{eqn:potentialdensity}
U = \cT(\phi) (1+\Psi_p) + Q \Phi \,.
\end{equation}
The pressure (force density) on the brane in the specified background is $P = - \nabla U$. To compute the pressure exerted on the brane by another brane, we calculate the background fields sourced by the second brane.

The background field action is
\begin{equation}
S_{\text{fields}} = \int d^d x \sqrt{-g} \left[ \frac{1}{2 \kappa_d^2} R - \frac{1}{2} G_{\phi\phi} (\nabla \phi)^2 - \frac{1}{2e^2} |F_{p+1}|^2\right].
\end{equation}
where $F_{p+1} = d A_p$ and $|F_{p+1}|^2 := F_{p+1} \cdot F_{p+1} = \frac{1}{(p+1)!} F_{M_1 \ldots M_{p+1}} F^{M_1 \ldots M_{p+1}}$. Varying $S = S_{\text{fields}} + S_p$ with respect to the background fields, we obtain the field equations
\begin{align}
  \frac{1}{2 \kappa_d^2} \biggl(R^{M N}-\frac{1}{2} g^{M N} R \biggr)  + (\ldots) &= - \frac{1}{2} \int d^p \xi \sqrt{-\tilde{g}}\, \frac{\delta^{(d)}(x - x (\xi))}{\sqrt{-g}} \cT(\phi) \tilde{g}^{a b} \partial_{a} x^{M} \partial_{b} x^N \,, \nonumber \\
  G_{\phi \phi} \nabla^2 \phi &=  \int d^p \xi \sqrt{-\tilde{g}}\, \frac{\delta^{(d)}(x - x (\xi))}{\sqrt{-g}} \cT'(\phi)\,, \nonumber \\
  \frac{1}{e^2} \nabla_{M_0} F^{M_0 \ldots M_p} &= Q \int d^p\xi \frac{\delta^{(d)}(x - x(\xi))}{\sqrt{-g}} \varepsilon^{a_1 \ldots a_p} \partial_{a_1} x^{M_1} \cdots \partial_{a_p} x^{M_p} \, ,
\end{align}
where $\varepsilon^{0 \cdots (p-1)} = +1$ is the brane Levi-Civita symbol and we omit terms of the Einstein equations that are quadratic order in the background fields $\nabla \phi$ and $F_{p+1}$.

Taking the brane to be stationary at $x^m = 0$, fixing Lorenz gauge $\nabla_{M_1} A^{M_1 \ldots M_p} = 0$, and linearizing about a fixed background $g_{M N} = \eta_{M N} + h_{M N}$ and $\phi = \phi_0 + \delta \phi$, we obtain
\begin{align}
  - \frac{1}{2} \partial^2 \bar{h}^{\mu \nu} &= -\kappa_d^2\, 
  \delta^{(d - p)} (x^m) \cT(\phi_0) \eta^{\mu \nu} \,, \nonumber \\
  G_{\phi \phi} \partial^2 \phi &=  \delta^{(d - p)}(x^m) \cT'(\phi_0) \,, &
  \frac{1}{e^2} \nabla^2 \Phi &= -Q \delta^{(d - p)} (x^m) \,,
\end{align}
where $\Phi = A_{0 \ldots (p-1)} = -A^{0 \ldots (p-1)}$, $\bar{h}_{M N} \df h_{M N} - \frac{1}{2} \eta_{M N} h$, and the remaining components of $A_{M_1 \ldots M_p}$ and $\bar{h}^{M N}$ besides those shown are not sourced by the brane.

Using (\ref{eqn:CoulombDdims}), we read off the solution,
\begin{align} \label{eqn:linearizedBraneFields}
\bar{h}_{\mu \nu} &= - \frac{2 \kappa_d^2 \cT(\phi_0)}{(d-p-2)V_{d-p-1}} \cdot \frac{\eta_{\mu \nu}}{r^{d-p-2}} \,, \nonumber \\
\phi &= \phi_0 - \frac{G_{\phi \phi}^{-1} \cT'(\phi_0)}{(d-p-2)V_{d-p-1}} \cdot \frac{1}{r^{d-p-2}} \,, &
\Phi &= \frac{e^2 Q}{(d-p-2)V_{d-p-1}} \cdot \frac{1}{r^{d-p-2}} \,.
\end{align}
Using $h_{M N} = \bar{h}_{M N} - \frac{1}{d-2} \eta_{M N} \bar{h}$, we obtain the Newtonian $p$-brane potential
\begin{equation}
\Psi_p = \frac{1}{2} \eta^{\mu \nu} h_{\mu \nu} = \frac{d-p-2}{2 (d-2)} \bar{h} = - \frac{p}{d-2} \cdot \frac{\kappa_d^2 \cT(\phi_0)}{V_{d-p-1} r^{d-p-2}} \,.
\end{equation}
Thus, applying~(\ref{eqn:potentialdensity}) and $P = -\nabla U$, we find the pressure on a brane exerted by a distant, parallel brane:
\begin{align}
P_{12} &= \frac{\mathcal{P}_{12}}{V_{d-p-1} r^{d-p-1}} \,, & \mathcal{P}_{12} = e^2 Q_1 Q_2 - G_{\phi \phi}^{-1} \cT'_1(\phi) \cT'_2(\phi) - \frac{p (d-p-2)}{d-2} \kappa_d^2 \cT_1(\phi) \cT_2(\phi) \,.
\end{align}
This is easily generalized to multiple gauge fields and moduli per \eqref{eqn:pformaction}:
\begin{equation} \label{eqn:boostInvariantPressure}
\mathcal{P}_{12} = f^{a b} Q_{1a} Q_{2b} - G^{i j} \partial_i \cT_1 \partial_j \cT_2 - \frac{p (d-p-2)}{d-2} \kappa_d^2 \cT_1 \cT_2 \,,
\end{equation}
where, as always, the moduli partial derivatives are taken with the $d$-dimensional Planck scale held fixed.

\subsection{General branes} \label{sec:generalbranes}

As seen above, a boost-invariant brane at rest in a flat background has the stress tensor
\begin{equation}
T_{\mu \nu} = \delta^{(d-p)}(x^m)\, \diag(\cT, - \cT, \ldots - \cT) \,,
\end{equation}
where $\cT$ is both the brane energy density and tension. This stress tensor is invariant under boosts parallel to the brane. On the other hand, subextremal black branes have a more general non-boost-invariant stress tensor
\begin{equation} \label{eqn:blackBraneStress}
T_{\mu \nu} = \delta^{(d-p)}(x^m)\, \diag(\mathcal{M}, -\mathcal{T}, \ldots, - \mathcal{T}) \,,
\end{equation}
in their rest frame, where the brane energy density $\mathcal{M}$ is no longer equal to the tension $\mathcal{T}$. Note that the null energy condition implies $\mathcal{M} \ge \mathcal{T}$; in~\secref{subsec:BHthermo} we show that this is satisfied for black branes and saturated if and only if the brane is quasiextremal.

To calculate the force between two non-boost-invariant branes we should in principle write down a probe action for each brane and proceed as above. The stress tensor for a black brane is that of a perfect fluid confined to the worldvolume with density $\rho = \mathcal{M}$ and pressure $p = -\mathcal{T}$. Thus, we might be inclined to write down a worldvolume action of the perfect fluid form, see, e.g.,~\cite{Brown:1992kc}. Whether this is the ``correct'' action depends on the physics of the black brane in question. Indeed, even assuming a perfect fluid worldvolume action, the brane dynamics will depend on an a priori unknown equation of state. Thus, while the Dirac action was essentially unique due to the assumption of worldvolume diffeomorphism invariance, the action for a black brane is much less constrained.

This is a serious obstacle, but fortunately we are only interested in the long range forces between widely separated branes. The stress tensor~(\ref{eqn:blackBraneStress}) is sufficient to determine the long-range gravitational fields sourced by the brane, hence it is also sufficient to determine the long range gravitational force (mediated by these fields) between two such branes.

It is convenient to generalize~(\ref{eqn:blackBraneStress}) to an arbitrary Lorentz frame:
\begin{equation} \label{eqn:blackBraneStressGen}
T_{\mu \nu} = \delta^{(d-p)}(x^m)\, \cT_{\mu \nu} \,,
\end{equation}
where $\cT_{\mu \nu}$ is the ``covariant tension,'' with $\cT_{\mu \nu} = \diag(\mathcal{M}, - \mathcal{T}, \ldots, - \mathcal{T})$ in the rest frame of the black brane. Note that the general covariant form of (\ref{eqn:blackBraneStressGen}) is
\begin{align}
T^{M N}(x) &= \int d^p \xi \sqrt{-\tilde{g}}\, \frac{\delta^{(d)}(x - x (\xi))}{\sqrt{-g}} \cT^{a b}(\xi) \partial_{a} x^{M} \partial_{b} x^N \,, 
 \end{align}
where the covariant tension $\cT^{a b}(\xi)$ is a worldvolume tensor and the integral defines the appropriate covariant delta function.

In general, unlike the boost-invariant case, $\mathcal{M} = \mathcal{M}(x)$ and $\mathcal{T} = \mathcal{T}(x)$ need not be constant along the brane worldvolume. In fact, for a generic brane they are dynamical quantities and can evolve with time, propagate disturbances, etc.. For simplicity, we will consider static, translation invariant black branes, so that $\mathcal{M}$ and $\mathcal{T}$ are constant along the worldvolume. However, unlike the mass of a particle or the tension of a boost-invariant brane (which can be thought of as a brane-localized cosmological constant), it is important not to confuse $\mathcal{M}$ and $\mathcal{T}$ with ``constants of nature'': taking such a brane and stretching it will in general change both $\mathcal{M}$ and $\mathcal{T}$.

 Solving the linearized Einstein equations, we obtain the gravitational field far from the brane
\begin{equation}
\bar{h}_{\mu \nu} = \frac{2 \kappa_d^2}{(d-p-2)V_{d-p-1}} \cdot \frac{\cT_{\mu \nu}}{r^{d-p-2}} \,. \label{eqn:farfield}
\end{equation}
Since the long-range gravitational field is linear in $\cT^{\mu \nu}$, the long-range force between two parallel branes must be bilinear in $\cT_1^{\mu \nu}$ and $\cT_2^{\mu \nu}$ as well as Poincar\'e covariant along their parallel worldvolumes. This fixes the general ansatz:
\begin{align} \label{eqn:branepressureAnsatz}
P_{12}^{\text{(grav)}} &= \frac{\mathcal{P}_{12}^{\text{(grav)}}}{V_{d-p-1} r^{d-p-1}} \,, & \mathcal{P}_{12}^{\text{(grav)}} &= -\kappa_d^2 (A\, \cT_1^{\mu \nu} \cT_{2\, \mu \nu} + B\, \cT_{1\; \mu}^{\mu} \cT_{2\; \nu}^{\nu} ) \,,
\end{align}
for coefficients $A$ and $B$ to be determined. Taking one of the branes to be boost invariant, $\cT_{\mu \nu} = - \cT \eta_{\mu \nu}$, and applying~(\ref{eqn:potentialdensity}) gives $A+p B = \frac{d-p-2}{d-2}$.

To fix the remaining linear combination of $A$ and $B$, we study the dynamics of a particular type of non-boost invariant brane. The easiest case to consider is that of a ``tensionless'' brane, i.e., one with a worldvolume action describing pressureless dust:
\begin{equation} \label{eqn:tensionlessbrane}
S= - \int d^{p-1}\xi d \tau \mathcal{M}_0(\xi) \sqrt{- g_{M N} \partial_\tau x^M \partial_\tau x^N} \,,
\end{equation}
where we omit couplings to gauge fields and moduli for the time being, and $\mathcal{M}_0(\xi)$ is a fixed (non-dynamical) positive function (the comoving density of the dust). This action is not invariant under general diffeomorphisms mixing $\xi^i$ and $\tau$, but it is invariant under $\tau \to \tilde{\tau} = \tilde{\tau}(\tau,\xi)$ as well as $\xi \to \tilde{\xi} = \tilde{\xi}(\xi)$ combined with $\mathcal{M}_0(\xi) \to \tilde{\mathcal{M}}_0(\tilde{\xi}) = \mathcal{M}_0(\xi) \det \frac{\partial \xi^m}{\partial \tilde{\xi}^n}$. Because of $\xi$ reparameterizations, the comoving density $\mathcal{M}_0(\xi)$ has no physical significance, and can be gauge-fixed to any positive value.

Varying the action with respect to the background metric, we find the covariant tension
$\cT^{a b} = \mathcal{M} u^a u^b$,
where
$u^\tau = \frac{1}{\sqrt{-\tilde{g}_{\tau \tau}}}$, $u^i = 0$ is the covariant velocity of the dust along the worldvolume and
\begin{equation} \label{eqn:invmassdensity}
\mathcal{M}(\xi,\tau) = \mathcal{M}_0(\xi) \frac{\sqrt{-\tilde{g}_{\tau \tau}}}{\sqrt{-\tilde{g}}}
\end{equation}
is the invariant mass density. Unlike $\mathcal{M}_0(\xi)$, the dynamical quantity $\mathcal{M}(\xi,\tau)$ is reparameterization invariant and physical.

The action (\ref{eqn:tensionlessbrane}) has enough gauge freedom to allow the gauge choice $\tau = x^0$.  Linearizing the background and expanding in small fluctuations, we obtain the potential density $U_0 = \mathcal{M}_0 (1+\Psi)$ \emph{per comoving volume} by the same methods as before, where $\Psi = - h_{tt}/2$ is the usual Newtonian potential for point particles. This is a very reasonable result: the brane can be thought of as a sheet of particles at rest with no short-range interactions between them, and reacts to gravitational fields in the same way that each particle in the sheet reacts.

From this, we obtain the force per comoving volume $P_0 = - \nabla U_0$ on the tensionless brane due to another brane with covariant tension $\cT_{2\, \mu\nu}$:
\begin{equation}
P_0 = - \frac{\kappa_d^2 \mathcal{M}_0}{V_{d-p-1} r^{d-p-1}} \biggl(\cT_{2\, t t} - \frac{1}{d-2} \eta_{t t} \cT^{\mu}_{2\; \mu}\biggr) \,.
\end{equation}
As noted before, the comoving density $\mathcal{M}_0$ is not invariant under $\xi$ reparameterizations. The pressure (force per physical volume) is instead
\begin{equation}
P = - \frac{\kappa_d^2 \mathcal{M} }{V_{d-p-1} r^{d-p-1}} \biggl(\cT_{2\, t t} - \frac{1}{d-2} \eta_{t t} \cT^{\mu}_{2\; \mu}\biggr) \,,
\end{equation}
with the invariant mass density $\mathcal{M}$ replacing the comoving mass density. 

Comparing with~(\ref{eqn:branepressureAnsatz}) fixes $A=1$ and $B = - \frac{1}{d-2}$. This agrees with the result $A+p B = \frac{d-p-2}{d-2}$ obtained using boost-invariant branes, a non-trivial consistency check of the calculation.

We conclude that the gravitational pressure exerted on one brane by a distant parallel brane takes the general form
\begin{align} \label{eqn:branepressureResult}
P_{12}^{\text{(grav)}} &= \frac{\mathcal{P}_{12}^{\text{(grav)}}}{V_{d-p-1} r^{d-p-1}} \,, & \mathcal{P}_{12}^{\text{(grav)}} &= -\kappa_d^2 \biggl[\cT_1^{\mu \nu} \cT_{2\, \mu \nu} - \frac{1}{d-2} \cT_{1\; \mu}^{\mu} \cT_{2\; \nu}^{\nu} \biggr] \,.
\end{align}
This corresponds to a brane energy density whose variation is
\begin{equation}
\delta U = -\frac{1}{2} \cT^{M N} \delta g_{M N} 
\end{equation}
for small perturbations about a flat background $\delta g_{M N} = h_{M N}$. This form follows directly from the definition of the stress tensor and the ansatz~(\ref{eqn:blackBraneStressGen}), so the result~(\ref{eqn:branepressureResult}) is completely general.

To complete our calculation, we reintroduce couplings to gauge bosons and moduli. The general result is:
\begin{equation} \label{eqn:branepressure}
\mathcal{P}_{12} = \f^{a b} Q_{1a} Q_{2b} - G^{i j} \mu_{1i} \mu_{1j} - \kappa_d^2 \biggl[\cT_1^{\mu \nu} \cT_{2\, \mu \nu} - \frac{1}{d-2} \cT_{1\; \mu}^{\mu} \cT_{2\; \nu}^{\nu} \biggr] \,.
\end{equation}
The last term is the gravitational contribution, discussed extensively above, whereas the first term is the gauge field contribution, which follows from the coupling $-q \int A_p$ independent of the details of the rest of the action. The middle term is mediated by the moduli, with the ``scalar charge'' $\mu_i$ defined by the linearized backreaction
\begin{equation} \label{eqn:scalarchargedef}
\phi^i = \phi^i_\infty - \frac{1}{(d-p-2)V_{d-p-1}}\cdot\frac{G^{i j} \mu_j}{r^{d-p-2}} + \ldots ,
\end{equation}
up to terms that are subleading in the large $r$ limit, where $G^{i j}$ is the inverse of the metric on moduli space, as before. The long range forces can only depend on the long range fields,\footnote{This is a consequence of Newton's third law, $\vec{F}_{12} = - \vec{F}_{21}$. Since the force on 1 due to 2, $\vec{F}_{12}$, depends only on the long range fields of 2, Newton's third law implies that $\vec{F}_{21} = - \vec{F}_{12}$ depends only on the long range fields of 2 (as well as the long range fields of 1), and therefore the long range forces depend only on the long range fields.} and therefore the scalar contribution to the pressure~(\ref{eqn:branepressure}) between two branes must be bilinear in their scalar charges $\mu_{1i}$ and $\mu_{2j}$. Diffeomorphism invariance in the moduli space implies that only the combination $G^{i j} \mu_{1i} \mu_{2j}$ can appear, where the constant prefactor can be fixed by comparing with (\ref{eqn:linearizedBraneFields}) and (\ref{eqn:boostInvariantPressure}).

Note that the scalar charge $\mu_i$ generalizes $\partial_i \cT$ in the Dirac brane case, but is no longer defined as a moduli derivative of the mass density and/or tension. There can still be a relation between $\mu_i$ and a moduli derivative, however, as in, e.g.,~\eqref{eqn:blackbranescalarcharge} below.
\subsection{Perfect branes} \label{subsec:perfectbrane}

We now check our calculation by considering a more general class of black brane encompassing both the Dirac and tensionless cases. Consider the action:
\begin{equation} \label{eqn:perfectbrane}
S= - \int d^{p-1}\xi d \tau \sqrt{-\tilde{g}} \mathcal{M}\biggl(s_0(\xi) \frac{\sqrt{-\tilde{g}_{\tau\tau}}}{\sqrt{-\tilde{g}}}, \phi\biggr)- Q \int A_p \,,
\end{equation}
for some fixed equation of state $\mathcal{M} = \mathcal{M}(s,\phi)$, where $s$ is the entropy density of a perfect fluid on the brane worldvolume and $\mathcal{M}$ is the energy density in the fluid rest frame. As in~(\ref{eqn:tensionlessbrane}), we choose comoving coordinates $\xi^i$ along the brane, which are constant along fluid flow lines. Like $\mathcal{M}_0(\xi)$ in~(\ref{eqn:tensionlessbrane}), $s_0(\xi)$ is a gauge-dependent positive function, whereas
\begin{equation}
s(\xi, \tau) = s_0(\xi) \frac{\sqrt{-\tilde{g}_{\tau\tau}}}{\sqrt{-\tilde{g}}} 
\end{equation}
is the physical entropy density. The action~(\ref{eqn:perfectbrane}) is that of a general perfect fluid confined to the worldvolume, in the limit where the fluid has no additional conserved quantities (such as particle number) besides its entropy.\footnote{This action can be obtained from that of~\cite{Brown:1992kc}, section 5, by taking the limit of zero number density (with fixed entropy density) and choosing Lagrangian coordinates on the brane.} I will refer to branes of this class as ``perfect branes.''

Varying the action~(\ref{eqn:perfectbrane}) with respect to the background metric, we obtain the covariant tension
\begin{align} \label{eqn:blackbranetension}
\cT^{a b} &= \cM u^a u^b-\cT (\tilde{g}^{a b} + u^a u^b)\,, & \cT &= \cM-s \left.\frac{\partial \cM}{\partial s}\right|_{\phi^i}  \,, & u^a = \frac{1}{\sqrt{- \tilde{g}_{\tau \tau}}} \delta^{a}_\tau \,.
\end{align}
Since $\mathcal{M}$ is the internal energy density, $T \df \left.\partial \cM / \partial s\right|_{\phi^i}$ is the brane temperature, and the tension $\cT$ is the Helmholz free energy density:\footnote{When the cosmological constant is non-vanishing, interpreting it as thermodynamic pressure suggests that the mass of a black hole is its thermodynamic \emph{enthalpy}~\cite{Kastor:2009wy,Dolan:2010ha}, see also~\cite{Cvetic:2010jb}. In this case, it seems likely that the tension of a perfect brane will correspond to its Gibbs free energy density. However, checking this is beyond the scope of the present paper.}
\begin{equation} \label{eqn:braneconstitutive}
\cT = \cM - T s \,.
\end{equation}
More generally, since the brane tension performs work, any black brane must satisfy $\cM - T s \ge \cT$, an inequality saturated by perfect branes. Assuming that $T s \ge 0$, this implies the null energy condition $\cT \le \cM$.

Likewise, varying the action~(\ref{eqn:perfectbrane}) with respect to the background moduli, we obtain the scalar charge
\begin{equation} \label{eqn:blackbranescalarcharge}
\mu_i = \left.\frac{\partial \mathcal{M}}{\partial \phi^i}\right|_s \,.
\end{equation}
This formula can be understood as follows. In a flat background with vanishing gauge fields, the brane has an energy density $U = \mathcal{M}(\phi, Y)$ where $Y$ schematically represents the internal degrees of freedom of the brane. Varying with respect to the moduli while holding the internal degrees of freedom fixed, we obtain
\begin{equation} \label{eqn:Mvariation}
\delta U = \left.\partial_i \mathcal{M}\right|_Y \delta \phi^i \,.
\end{equation}
Thus, the brane is subject to a pressure $P = - \left.\partial_i \mathcal{M}\right|_Y \nabla \phi^i$. Comparing with~(\ref{eqn:branepressure}) and~(\ref{eqn:scalarchargedef}), we read off $\mu_i = \left.\partial_i \mathcal{M}\right|_Y$. Since the brane entropy density $s$ depends only on the internal degrees of freedom $Y$, and in the perfect case the latter are fixed by the former, $\left.\partial_i \mathcal{M}\right|_Y = \left.\partial_i \mathcal{M}\right|_s$, and we recover~\eqref{eqn:blackbranescalarcharge}. This agrees with (\ref{eqn:boostInvariantPressure}), where in the Dirac case $s = 0$ identically.

The black brane solutions that arise from the two-derivative effective actions we consider will turn out to satisfy~\eqref{eqn:braneconstitutive} and~\eqref{eqn:blackbranescalarcharge} (see \eqref{eqn:blackbraneconstitutive} and \eqref{eqn:scalarcharge2}), strongly suggesting that these black branes are likewise perfect.

To compute the force on a perfect brane in some background, we choose the gauge $\tau = x^0$, expand in small fluctuations, and linearize the background. After some calculation, one obtains the brane potential energy density
\begin{equation}
U = \mathcal{M}(s_0, \phi) - \frac{1}{2} h_{t t} \mathcal{M}(s_0, \phi) + \frac{1}{2} h^i_{\; i} \mathcal{T}(s_0, \phi) +Q \Phi = \mathcal{M}(s_0, \phi) - \frac{1}{2} \cT^{M N} h_{M N} +Q \Phi \,,
\end{equation}
for small uniform displacements in the transverse directions, where $s_0$ is the unperturbed brane entropy density. From this, we readily recover~(\ref{eqn:branepressure}) using the methods previously described.

\subsubsection{Dimensional reduction}

It is interesting to consider what happens to the long-range forces after dimensional reduction on a circle of radius $R$. This was analyzed in the case of Dirac branes in~\cite{Heidenreich:2019zkl}, with the result that the sign of the force between two parallel branes is unchanged. I now briefly describe how this works for perfect branes.

Suppose first that the brane is transverse to the compact circle. In this case, the mass density of the dimensionally reduced brane is the same as the original, $\mathcal{M}_d = \mathcal{M}_D$, where $D = d+1$ and $d$ are the spacetime dimensions before and after compactification, respectively. Since $\mathcal{M}$ does not depend on the radius, naively one might think the brane carries no radion charge, where the radion  $\rho \df \log(2 \pi R M_d)$ parameterizes the circle radius in units of the $d$-dimensional Planck scale $M_d^{d-2} = 1/\kappa_d^2$.
This is not the case because (as noted previously) derivatives involving dimensionful quantities implicitly hold the Planck scale fixed, and the $d$-dimensional Planck scale differs from the $D$-dimensional Planck scale by a radion-dependent factor, $M_d^{d-2} = 2\pi R M_D^{D-2}$. Both $\mathcal{M}$ and $s$ are dimensionful, giving two independent contributions to the radion charge.

We explicitly compute the radion charge as follows:
\begin{equation}
  \frac{\partial \mathcal{M}}{\partial \rho} \bigg|_{s, \phi}^{(d)}
  \df M_d^p  \frac{\partial}{\partial \rho}  \frac{\mathcal{M}}{M_d^p} \biggr|_{s / M_d^{p - 1}, \phi}
  =  \mathcal{M} \frac{M_d^p}{M_D^p} \frac{d}{d \rho}  \frac{M_D^p}{M_d^p}
  + M_D^p  \frac{\partial}{\partial \rho} \frac{\mathcal{M}}{M_D^p} \biggr|_{s / M_d^{p - 1}, \phi} .
\end{equation}
The derivative in the second term fixes $s / M_d^{p - 1}$, but $s / M_D^{p - 1} = \frac{M_d^{p - 1}}{M_D^{p - 1}}\cdot(s / M_d^{p - 1})$ depends on $\rho$, and we obtain
\begin{multline}
  M_D^p \frac{\partial}{\partial \rho} \frac{\mathcal{M}}{M_D^p} \biggr|_{s / M_d^{p - 1}, \phi} 
  = \biggl( \frac{s}{M_d^{p - 1}} \frac{d}{d \rho} \frac{M_d^{p - 1}}{M_D^{p - 1}} \biggr)
  M_D^p \frac{\partial}{\partial (s / M_D^{p - 1})} \frac{\mathcal{M}}{M_D^p} \biggr|_{\phi} \\
  = \biggl( \frac{M_D^{p - 1}}{M_d^{p - 1}} \frac{d}{d \rho} \frac{M_d^{p - 1}}{M_D^{p - 1}} \biggr) s \frac{\partial \mathcal{M}}{\partial s} \biggr|^{(D)}_{\phi}\,.
\end{multline}
To compute the radion derivatives, note that applying $M_d^{d - 2} = 2 \pi R M_D^{D - 2}$ to eliminate $R$ gives $\rho = (d-1) \log \frac{M_d}{M_D}$.
Thus,
\begin{equation} \label{eqn:radiondependence}
  \frac{\partial \mathcal{M}}{\partial \rho} \bigg|_{s, \phi}^{(d)} = -\frac{p}{d-1} \mathcal{M} + \frac{p-1}{d-1} s \frac{\partial \mathcal{M}}{\partial s} \biggr|^{(D)}_{\phi} .
\end{equation}
The first term arises because $\mathcal{M}$ has dimension $p$, whereas the second arises because $s$ has dimension $p-1$.

We can rexpress \eqref{eqn:radiondependence} in terms of the tension using~\eqref{eqn:blackbranetension}:
\begin{equation}
\frac{\partial \mathcal{M}}{\partial \rho} \bigg|_{s, \phi}^{(d)} = -\frac{p}{d-1} \mathcal{M} + \frac{p-1}{d-1} (\cM - \cT) = -\frac{1}{d-1} \cT^{\mu}_{\; \mu} .
\end{equation}
Using this simple result, it is straightforward to check that
\begin{equation}
  2 \pi R \biggl(G^{\rho \rho}  \biggl[ \frac{\partial \mathcal{M}}{\partial \rho}\biggr]^2 + \kappa_d^2 \biggl[\cT^{\mu \nu} \cT_{\mu \nu} - \frac{1}{d-2} \cT_{\; \mu}^{\mu} \cT_{\; \nu}^{\nu} \biggr] \biggr)
  = \kappa_D^2 \biggl[\cT^{\mu \nu} \cT_{\mu \nu} - \frac{1}{D-2} \cT_{\; \mu}^{\mu} \cT_{\; \nu}^{\nu} \biggr] \,,
\end{equation}
where $\kappa_D^2 = (2\pi R) \kappa_d^2$, $G^{\rho \rho} = \kappa_d^2  \frac{d - 1}{d - 2}$,\footnote{See, e.g.,~\cite{Heidenreich:2015nta}, where $\lambda^{(\text{there})} = -2\log(2\pi R M_D) = -2 \frac{d-2}{D-2} \rho$ (up to an additive constant).} and the effect of the radion coupling is to change $\frac{1}{d-2}$ to $\frac{1}{D-2}$ in the last term. Matching the other moduli and gauge forces, we obtain:
\begin{equation} \label{eqn:preservingP}
\mathcal{P}_d = \frac{1}{2 \pi R} \mathcal{P}_D \,,
\end{equation}
where $\cP$ is the pressure coefficient defined in~\eqref{eqn:branepressureResult}.
This is the same result as~\cite{Heidenreich:2019zkl}, generalized to perfect branes.

Suppose instead that the brane wraps the compact circle, so that a $(P-1)$-brane in $D$-dimensions produces a $(p-1)$-brane in $d$-dimensions for $P=p+1$. In this case the $d$-dimensional mass density is explicitly $R$-dependent, $\mathcal{M}_d = (2\pi R) \mathcal{M}_D$. By a similar calculation to before,
\begin{equation} \label{eqn:radiondependence2}
\frac{1}{2\pi R} \frac{\partial \mathcal{M}_d}{\partial \rho} \bigg|_{s_d, \phi}^{(d)} =\frac{d-p-2}{d-1} \mathcal{M}_D - \frac{d-p-1}{d-1} s_D \frac{\partial \mathcal{M}_D}{\partial s_D} \biggr|^{(D)}_{\phi} ,
\end{equation}
where $s_d = (2 \pi R) s_D$. In comparison with~\eqref{eqn:radiondependence}, there is an extra $\frac{d-2}{d-1}$ contribution to each term originating from the explicit factors of $R$ relating $\mathcal{M}_d$ with $\mathcal{M}_D$ and $s_d$ with $s_D$. Expressing this in terms of the tension we obtain:
\begin{equation} \label{eqn:radionchargeReducingP}
\frac{1}{2\pi R} \frac{\partial \mathcal{M}_d}{\partial \rho} \bigg|_{s_d, \phi}^{(d)}  = - \frac{\mathcal{M}_D + p \mathcal{T}_D}{d- 1} + \mathcal{T}_D = -\frac{1}{d-1} \cT_{D\; \mu}^{\mu} + \cT_D .
\end{equation}
Note that
\begin{equation} \label{eqn:stresstensorReducingP}
\begin{aligned}
\frac{1}{2\pi R} \cT_{d\; \mu}^{\mu} &= \cM_D + (p-1) \cT_D =  \cT_{D\; \mu}^{\mu} - \cT_D \,,  \\
\frac{1}{(2\pi R)^2} \cT_d^{\mu \nu} \cT_{d\; \mu \nu} &= \cM_D^2 + (p-1) \cT_D^2 = \cT_D^{\mu \nu} \cT_{D\; \mu \nu} - \cT_D^2 \,.
\end{aligned}
\end{equation}
Using these formulae, we find
\begin{equation}
  \frac{1}{2 \pi R} \biggl(G^{\rho \rho}  \biggl[ \frac{\partial \mathcal{M}_d}{\partial \rho}\biggr]^2 + \kappa_d^2 \biggl[\cT_d^{\mu \nu} \cT_{d\; \mu \nu} - \frac{1}{d-2} \cT_{d\; \mu}^{\mu} \cT_{d\; \nu}^{\nu} \biggr] \biggr)
  = \kappa_D^2 \biggl[\cT_D^{\mu \nu} \cT_{D\; \mu \nu} - \frac{1}{D-2} \cT_{D\; \mu}^{\mu} \cT_{D\; \nu}^{\nu} \biggr] \,,
\end{equation}
similar to before, where the extra $\cT_D$ and $\cT_D^2$ terms in~\eqref{eqn:radionchargeReducingP} and~\eqref{eqn:stresstensorReducingP} conspire to cancel. Thus, we obtain
\begin{equation} \label{eqn:reducingP}
\mathcal{P}_d = (2 \pi R) \mathcal{P}_D \,,
\end{equation}
again in agreement with~\cite{Heidenreich:2019zkl}.

In fact, the results~\eqref{eqn:preservingP} and~\eqref{eqn:reducingP} hold more generally for sufficiently large $R$. In the large $R$ limit we can understand the effect of dimensional reduction on the long-range forces by considering the long-range fields only, without knowledge of the brane action. Since any choice of $\cT_{\mu \nu}$, $\mu_i$, and $q_a$ can be realized by a perfect brane, \eqref{eqn:preservingP} and~\eqref{eqn:reducingP} apply to arbitrary uniform, parallel branes.

\section{Black hole and black brane solutions} \label{sec:BH}

Armed with a thorough understanding of long range forces, we now consider the details of spherically symmetric black hole and black brane solutions to the two-derivative effective action, subject to the assumptions discussed in~\secref{sec:setup}. By a suitable choice of gauge, we characterize quasiextremal and non-extremal solutions and show that the former have vanishing self-force, whereas the latter are self-attractive. Some illustrative examples from the literature are reviewed in appendix~\ref{app:examples}

\subsection{Equations of motion}

As argued in section~\secref{subsec:branesetup}, for a spherically symmetric $(p-1)$-brane with only $p$-form charge, we can truncate to the action~\eqref{eqn:pformaction}. The corresponding equations of motion are
\begin{align}
 d (\f_{a b} (\phi) \ast F^b) &= 0 \qquad, \qquad \nabla^2 \phi^i +
  \Gamma^i_{\; j k} (\phi) \nabla \phi^j \cdot \nabla \phi^k = \frac{1}{2}
  G^{i j} (\phi) \f_{a b, j} (\phi) F^a \cdot F^b \;,  \nonumber \\
  R_{M N} - \frac{1}{2} g_{M N} R &= \kappa_d^2 G_{i j} (\phi) 
  \nabla_M \phi^i \circ \nabla_N \phi^j + \kappa_d^2 \f_{a b} (\phi) F^a_M \circ
  F^b_N  , 
\end{align}
where\footnote{See~\eqref{eqn:dotdefn} for the definition of $\omega\cdot\chi$.}
\begin{align}
\omega_M \circ \chi_N &\df \omega_{(M} \cdot \chi_{N)} - \frac{1}{2} g_{M N}\, \omega\cdot\chi \,, &  \omega_M^{(p+1)} \cdot \chi_N^{(p+1)} &\df \frac{1}{p !} \omega_{M N_1 \ldots N_p}
  \chi_N^{\;  \; N_1 \ldots N_p} \;, \label{eqn:circdef}
\end{align}
$G^{i j} (\phi)$ is the inverse of the metric on moduli space $G_{i j} (\phi)$, and
\begin{equation}
  \Gamma^i_{\; j k} (\phi) \df \frac{1}{2} G^{i l} (\phi)  [G_{l j, k}
  (\phi) + G_{l k, j} (\phi) - G_{j k, l}(\phi)]
\end{equation}
are the coefficients of the corresponding metric connection.

We consider the general static spherically symmetric black brane ansatz
\begin{align}
  d s^2 &= - e^{2 \psi_t (r)} d t^2 + e^{2 \psi_y (r)} d y^2
  + e^{2 \psi_r (r)} d r^2 + r^2 e^{2 \psi_{\Omega} (r)} d \Omega^2_{d - p -
  1} \,, \nonumber \\
  \phi^i &= \phi^i (r) \;, \qquad A_p^a = \Phi^a (r) d t \wedge d y^1 \wedge
  \ldots \wedge d y^{p-1} \,,
\end{align}
where $(t, y^m)$ are the directions parallel to the black brane and $dy^2 \df \delta_{m n} d y^m d y^n$ is the Euclidean metric in the $y$ directions. The charge of the black brane is measured by the integral
\begin{equation}
  Q_a = (-1)^p \oint_{S^{d - p - 1}} \!\!\!\!\!\! \f_{a b} (\phi) \ast F^b \, , \qquad \text{which implies} \quad \f_{a b} (\phi) \ast F^b = (-1)^p Q_a \frac{\omega_{d - p - 1}}{V_{d-p-1}}
\end{equation}
using spherical symmetry,
where $\omega_{d - p - 1}$ is the volume-form for the unit metric on $S^{d - p- 1}$. 
From this we obtain,
\begin{align}
 F^a &= - \frac{\f^{a b} (\phi) Q_b}{V_{d - p - 1}}  \frac{e^{\psi_t +
  \psi_r + (p - 1) \psi_y - (d - p - 1) \psi_{\Omega}}}{r^{d - p - 1}} d r
  \wedge d t \wedge d y^1 \wedge \ldots \wedge d y^{p - 1} \,, \nonumber \\
  \Phi^a (r) &= \frac{1}{V_{d - p - 1}}  \int_r^{\infty} \f^{a b} (\phi) Q_b 
  \frac{e^{\psi_t + \psi_r + (p - 1) \psi_y - (d - p - 1) \psi_{\Omega}}}{r^{d
  - p - 1}} d r \,, \label{eqn:PhiInt0}
\end{align}
where $\f^{a b}(\phi)$ is the inverse of gauge kinetic matrix $\f_{a b}(\phi)$. We can eliminate the gauge field from the remaining equations of motion in favor of $Q_a$. In particular,
\begin{gather}
\f_{a b} F^a \!\!\cdot\!\! F^b = -e^{-2 \psi_t} \f_{a b} F^a_t  \!\!\cdot\!\! F^b_t = e^{-2 \psi_r} \f_{a b} F^a_r  \!\!\cdot\!\!  F^b_r = e^{-2 \psi_y} \f_{a b} F^a_m  \!\!\cdot\!\!  F^b_m = - \frac{Q^2(\phi)}{V_{d - p - 1}^2} 
  \frac{e^{- 2 (d - p - 1) \psi_{\Omega}}}{r^{2 (d - p - 1)}} \,, \nonumber \\
  \f_{a b} F^a_{m} \!\!\cdot\!\!  F^b_{n} = 0 \quad (m\ne n), \qquad \f_{a b} F^a_{\alpha} \!\!\cdot\!\!  F^b_{\beta} = 0 \,,
\end{gather}
where $\alpha, \beta$ index the angular directions and $Q^2(\phi) \df \f^{a b}(\phi) Q_a Q_b$.

Thus, writing the components of the Ricci tensor as
\begin{equation}
  R_{t t} = \frac{g_{t t}}{g_{r r}} \mathcal{R}_t \;, \qquad R_{r r}
  =\mathcal{R}_r \;, \qquad R_{i j} = \frac{g_{i j}}{g_{r r}} \mathcal{R}_y
  \;, \qquad R_{\alpha \beta} = \frac{g_{\alpha \beta}}{g_{r r}} \mathcal{R}_{\Omega} \,,
\end{equation}
we obtain the Einstein equations
\begin{equation}
  \mathcal{R}_t =\mathcal{R}_y =\mathcal{R}_r - \kappa_d^2 G_{i j}
  \frac{d \phi^i}{d r} \frac{d \phi^j}{d r} = -  \frac{d - p - 2}{d - 2} \frac{\kappa_d^2 Q^2}{V_{d - p - 1}^2}
  \frac{e^{2 \psi_r - 2 (d - p - 1) \psi_{\Omega}}}{r^{2
  (d - p - 1)}} = - \frac{d - p - 2}{p} \mathcal{R}_{\Omega} \, .
\end{equation}
Likewise, the scalar equations of motion become
\begin{multline}
  \frac{d^2 \phi^i}{dr^2} + \biggl[ \psi_t' + (p - 1) \psi_y' - \psi_r' + (d - p -
  1) \Bigl(\psi_{\Omega}'+\frac{1}{r}\Bigr) \biggr] \frac{d \phi^i}{dr} +
  \Gamma^i_{\; j k} (\phi) \frac{d\phi^j}{dr} \frac{d\phi^k}{dr} \\=
\frac{1}{2 V_{d - p - 1}^2} G^{i j} Q^2_{, j}
  \frac{e^{2 \psi_r - 2 (d - p - 1) \psi_{\Omega}}}{r^{2 (d - p - 1)}} \;,   
\end{multline}
where $Q^2_{,j} (\phi) = \partial_j Q^2(\phi)$.
An explicit calculation gives
\begin{align}
  \mathcal{R}_t &= \psi_r' \psi_t' - \psi_t'' - (\psi_t')^2 - (p - 1) \psi_t'
  \psi_y' - (d - p - 1) \psi_t'  \left( \psi_{\Omega}' + \frac{1}{r} \right) , \nonumber \\
  \mathcal{R}_r &= \psi_t' \psi_r' - (\psi_t')^2 - \psi_t'' + (p - 1) 
  [\psi_y' \psi_r' - (\psi_y')^2 - \psi_y''] \nonumber \\
  &\noeq + (d - p - 1)  \left[
  \psi_r' \psi_{\Omega}' - (\psi_{\Omega}')^2 - \psi_{\Omega}'' +
  \frac{\psi_r' - 2 \psi'_{\Omega}}{r} \right] , \nonumber \\
  \mathcal{R}_y &= \psi_r' \psi_y' - \psi_y'' - (p - 1)  (\psi_y')^2 -
  \psi_y' \psi_t' - (d - p - 1) \psi_y'  \left( \psi_{\Omega}' + \frac{1}{r} \right) , \nonumber \\
  \mathcal{R}_{\Omega} &= \frac{1}{r^2} - \psi_{\Omega}'' + \left(
  \psi_{\Omega}' + \frac{1}{r} \right)  [\psi_r' - \psi_t' - (p - 1) \psi_y']
  - (d - p - 1) \left( \psi_{\Omega}' + \frac{1}{r} \right)^2 \nonumber \\
  &\noeq + \frac{d - p - 2}{r^2} e^{2 (\psi_r - \psi_{\Omega})} \; ,  
\end{align}
in agreement with~\cite{duff:1996hp}.\footnote{There is a typo in the expression for $R_{tt}$ in~\cite{duff:1996hp}: $(v')^2_{\mathrm{there}}$ should actually be
$(u')^2_{\mathrm{there}}$.}

\subsection{Solutions with a smooth horizon} \label{subsec:BHsolns}

The above equations are invariant under radial diffeomorphisms. We choose the gauge:
\begin{equation}
  \psi_t + \psi_r + (p - 1) \psi_y + (d - p - 3) \psi_{\Omega} = 0 \,,
\end{equation}
which can be parameterized as:
\begin{equation} \label{eqn:lambdagauge}
  d s^2 = e^{\frac{2}{p} \psi}  \left[ - f (r) e^{- 2 \frac{p - 1}{p} \lambda}
  d t^2 + e^{\frac{2}{p} \lambda} d y^2 \right] + e^{-
  \frac{2}{d - p - 2} \psi}  \left[ \frac{d r^2}{f (r)} + r^2 d \Omega^2_{d -
  p - 1} \right] \;,
\end{equation}
for some $\psi (r), \lambda(r)$ and $f(r)$ to be determined. Note that $\lambda(r)$ disappears from the ansatz in the black hole case, $p=1$, since the transverse directions $y^m$ are absent.

To check for a smooth horizon, it is convenient to rewrite this as
\begin{equation} \label{eqn:RYansatz}
  d s^2 = - \frac{F (\rho) d t^2}{R (\rho)^{2 (d - p - 2)} Y (\rho)^{2 (p - 1)}}  + Y (\rho)^2 dy^2 + R (\rho)^2  \left[ \frac{d
  \rho^2}{(d - p - 2)^2 F (\rho)} + d \Omega^2_{d - p - 1} \right] \;,
\end{equation}
where
\begin{equation}
  \rho \df r^{d - p - 2}, \qquad Y (\rho) = e^{\frac{1}{p}  (\psi +
  \lambda)}, \qquad R (\rho) = r e^{- \frac{1}{d - p - 2} \psi}, \qquad F
  (\rho) = r^{2 (d - p - 2)} f (r) . \label{eqn:rhoYRFdef}
\end{equation}
Putting~\eqref{eqn:RYansatz} into ingoing Eddington-Finkelstein coordinates, we
obtain:
\begin{equation}
  d s^2 = - \frac{F (\rho) d v^2}{R^{2 (d - p - 2)} Y^{2 (p - 1)}} + \frac{2 d
  v d \rho}{(d - p - 2) R^{d - p - 3} Y^{p - 1}} + Y^2 d y^2 + R^2 d \Omega^2_{d - p - 1} .
\end{equation}
In this form, it is clear that to have a smooth horizon, $R (\rho)$ and (for $p>1$) $Y
(\rho)$ must remain finite while $F (\rho) \rightarrow 0$ as $\rho \rightarrow
\rho_h$ for finite $\rho_h$. There is a residual gauge symmetry $\rho
\rightarrow \rho + \varepsilon$ for constant $\varepsilon$ that we will fix below.

Returning to the ansatz~\eqref{eqn:lambdagauge}, we compute:
\begin{equation}
  -\mathcal{R}_t - (p - 1) \mathcal{R}_y - (d - p - 2) \mathcal{R}_{\Omega} =
  \frac{f''}{2 f} + \frac{3 d - 3 p - 5}{2 r}  \frac{f'}{f} + \frac{(d - p -
  2)^2}{r^2}  \frac{f - 1}{f} .
\end{equation}
Thus, the Einstein equations imply:
\begin{equation}
  f'' + \frac{3 d - 3 p - 5}{r} f' + 2 \frac{(d - p - 2)^2}{r^2}  (f - 1) = 0,
\end{equation}
which has the solution:
\begin{equation}
  f (r) = 1 + \frac{A}{r^{d - p - 2}} + \frac{B}{r^{2 (d - p - 2)}}\,, \qquad \text{corresponding to} \quad F (\rho) = \rho^2 + A \rho + B \,. \label{eqn:fgensoln}
\end{equation}
$F (\rho)$ must have a zero for finite $\rho$ in order to have a horizon.
Thus, we require $A^2 \geqslant 4 B$, and $F (\rho)$ can be factored:
\begin{equation}
  F (\rho) = (\rho - \rho_-) (\rho - \rho_+),
\end{equation}
for $\rho_- \leqslant \rho_+$. We use the residual gauge symmetry to set
$\rho_- = 0$, so that $F (\rho) = \rho (\rho - \rho_h)$ for $\rho_h \geqslant
0$. In this gauge, we have
\begin{equation}
  f (r) = 1 - \frac{r_h^{d - p - 2}}{r^{d - p - 2}},
\end{equation}
for $r_h \geqslant 0$. (We later show that $r_h = 0$ if and only if the solution is quasiextremal.)

For $p>1$, we find:
\begin{equation}
  \mathcal{R}_t -\mathcal{R}_y = \lambda'' + \left[ \frac{f'}{f} + \frac{d - p
  - 1}{r} \right] \lambda' - \left( \frac{f''}{2 f} + \frac{f'}{2 f}  \frac{d
  - p - 1}{r} \right),
\end{equation}
but the term in parentheses vanishes in the gauge $\rho_- = 0$, so the Einstein equations imply
\begin{equation}
  \lambda'' + \left[ \frac{f'}{f} + \frac{d - p - 1}{r} \right] \lambda' = 0 .
\end{equation}
The solution for $r_h > 0$ is
\begin{equation}
  \lambda = C \log f + D,
\end{equation}
for constants $C$ and $D$, where $D = 0$ to have $\lambda \rightarrow 0$ as $r
\rightarrow \infty$. However, since $R (\rho)$ and $Y (\rho)$ must be finite
at the horizon, $e^{\frac{1}{p} \lambda}$ must also be finite at the horizon.
Therefore $C = 0$, implying that $\lambda = 0$.

If $r_h = 0$ then $f = 1$, and the solution is instead:
\begin{equation}
  \lambda = \frac{C}{r^{d - p - 2}} + D,
\end{equation}
with $D = 0$ to preserve asymptotic flatness, as before. Consider:
\begin{equation}
  R^{d - p - 2} Y^p = r^{d - p - 2} e^{\lambda} = r^{d - p - 2} \exp \biggl(
  \frac{C}{r^{d - p - 2}} \biggr) .
\end{equation}
Regardless of the value of $C$, this is not finite as $r \rightarrow 0$,
so a smooth horizon is impossible for $r_h = 0$ when $p > 1$.

Thus, in our chosen gauge all solutions with smooth horizons take the form
\begin{equation}
  d s^2 = e^{\frac{2}{p} \psi}  [- f (r) d t^2 + d y^2] +
  e^{- \frac{2}{d - p - 2} \psi}  \left[ \frac{d r^2}{f (r)} + r^2 d
  \Omega^2_{d - p - 1} \right] \;, \label{eqn:finalmetricansatz}
\end{equation}
with $f (r) = 1 - \frac{r_h^{d - p - 2}}{r^{d - p - 2}}$, $r_h \geqslant 0$,
and $\psi (r)$ to be determined. Note that for $p>1$ boost-invariance is restored as $r_h \to 0$, though a smooth horizon is lost in this limit.

With the above gauge choice and solution for $f$ and $\lambda$, the remaining
non-vanishing components of the Ricci tensor are:
\begin{align}
  -\mathcal{R}_t - (p - 1) \mathcal{R}_y &= \psi'' + \biggl[ \frac{f'}{f} +
  \frac{d - p - 1}{r} \biggr] \psi'\,, \nonumber \\
  \mathcal{R}_t + (p - 1) \mathcal{R}_y + (d - p - 1) \mathcal{R}_{\Omega}
  -\mathcal{R}_r &= \frac{d - 2}{p (d - p - 2)} \psi'  \biggl( \psi' +
  \frac{f'}{f} \biggr) \,.
\end{align}
Thus, the scalar equations of motion and the remaining Einstein equations are
\begin{subequations}  \label{eqn:simpEOM}
\begin{align}
  \frac{1}{r^{d - p - 1} f} d_r [r^{d - p - 1} f d_r \phi^i] + \Gamma^i_{\; j
  k} (\phi) d_r \phi^j d_r \phi^k &= \frac{1}{2} \mathcal{G}^{i j} (\phi)
  \mathcal{Q}^2_{, j} (\phi)  \frac{e^{2 \psi}}{r^{2 (d - p - 1)} f} \,,  \label{eqn:psirEOM} \\
  \frac{1}{r^{d - p - 1} f} d_r [r^{d - p - 1} f \psi'] &= \mathcal{Q}^2
  (\phi)  \frac{e^{2 \psi}}{r^{2 (d - p - 1)} f} \,, \label{eqn:phirEOM} \\
  \psi'  \left( \psi' + \frac{f'}{f} \right) +\mathcal{G}_{i j} (\phi)
  d_r \phi^i d_r \phi^j &= \mathcal{Q}^2 (\phi)  \frac{e^{2
  \psi}}{r^{2 (d - p - 1)} f} \,, \label{eqn:rCons}
\end{align}
\end{subequations}
where
\begin{equation} \label{eqn:BHpotential}
  \mathcal{Q}^2 (\phi) \df \frac{\xi \kappa_d^2}{V_{d - p - 1}^2} Q^2(\phi) \,, \qquad \mathcal{G}_{i j}(\phi) = \xi \kappa_d^2 G_{i j}(\phi) \,, \qquad \xi \df \frac{p (d - p - 2)}{d - 2} \,.
\end{equation}
and $d_r$ is a shorthand for $\frac{d}{dr}$. Note that~\eqref{eqn:rCons} is a constraint equation (containing no second derivatives); its $r$ derivative vanishes upon imposing~\eqref{eqn:psirEOM}, \eqref{eqn:phirEOM}, as required for consistency.

Defining the inverse radial variable
\begin{equation}
  z \df \frac{1}{(d - p - 2) r^{d - p - 2}} \,,
\end{equation}
the equations~\eqref{eqn:simpEOM} become still simpler:
\begin{subequations}  \label{eqn:zEOM}
\begin{align}
  d_z [f \dot{\psi}] &=  \mathcal{Q}^2 (\phi) e^{2 \psi} \,, \label{eqn:psizEOM} \\
   d_z [f
  \dot{\phi}^i] + f \Gamma^i_{\; j k} (\phi)  \dot{\phi}^j  \dot{\phi}^k &=
  \frac{1}{2} \mathcal{G}^{i j} (\phi) \mathcal{Q}^2_{, j} (\phi) e^{2
  \psi}, \label{eqn:phizEOM}  \\
  \dot{\psi} (f \dot{\psi} + \dot{f}) +  f\mathcal{G}_{i j}
  (\phi) \dot{\phi}^i  \dot{\phi}^j &=  \mathcal{Q}^2 (\phi) e^{2 \psi}, \label{eqn:zCons}
\end{align}
\end{subequations}
where dots denote derivatives with respect to $z$. Here $z = 0$
corresponds to spatial infinity, $r = \infty$, and $z$ increases as $r$ decreases with the
event horizon at $z_h \df [(d - p - 2) r^{d - p - 2}_h]^{- 1}$. Thus, $f =
1 - \frac{z}{z_h}$, and $\dot{f} = - \frac{1}{z_h}$ is a constant.

Note that the equations~\eqref{eqn:zEOM} are invariant under
\begin{align}
z &\to z' = a z + b, & z_h &\to z_h' = a z_h + b, & e^{2 \psi} &\to e^{2 \psi'}=\frac{z_h}{a (a z_h+b)} e^{2 \psi}, \label{eqn:zLinearSymm}
\end{align}
which implies $f \to f'=\frac{a z_h}{a z_h + b} f$. Moreover, note that~\eqref{eqn:psizEOM} and~\eqref{eqn:zCons} together imply the useful relation
\begin{equation}
  \ddot{\psi} = \dot{\psi}^2 +  \mathcal{G}_{i j} (\phi) \dot{\phi}^i \dot{\phi}^j \,. \label{eqn:psiNoQ}
\end{equation}

\subsubsection{General properties of solutions with smooth horizons}

While the solutions to~\eqref{eqn:zEOM} will depend on the charge function $\mathcal{Q}^2(\phi)$ as well as on the metric on moduli space $\mathcal{G}_{i j}(\phi)$, we can understand their general properties without knowing these functions. For $N$ moduli, there are $N+1$ dynamic variables $\psi$ and $\phi^i$ satisfying second-order equations of motion, with a single constraint equation. Thus, a general solution depends on $2N+1$ free parameters. However, the requirement of a smooth horizon reduces the number of free parameters, as follows.

If $r_h > 0$, then $\psi$ and $\phi^i$ must remain finite at the horizon. With this assumption, the equations of motion degenerate to first order equations at the horizon, giving
\begin{equation}
  - \frac{1}{z_h}  \dot{\psi} (z_h) =  \mathcal{Q}^2 (\phi_h) e^{2 \psi_h}
  \;, \qquad - \frac{1}{z_h}  \dot{\phi}^i (z_h) = \frac{1}{2} \mathcal{G}^{i
  j} (\phi_h) \mathcal{Q}^2_{, j} (\phi_h) e^{2 \psi_h} \;,
  \label{eqn:smoothhorizon}
\end{equation}
assuming the solution remains regular there. Higher $z$ derivatives can be
fixed by taking derivatives of the equations of motion and setting $z = z_h$.
Thus, for fixed charges an arbitrary solution for $\psi, \phi^i$ that is smooth at the horizon
has exactly $N + 1$ free parameters $\psi_h$ and $\phi_h^i$.\footnote{This argument implicitly assumes that $\cG_{i j}(\phi)$ and $\cQ^2(\phi)$ are analytic at $\phi = \phi_h$. If not, then $\psi(z)$ and $\phi^i(z)$ do not have convergent power series at $z=z_h$, and the number of integration constants cannot be determined in this way.} 
Note that asymptotic flatness requires $\psi=0$ at $z=0$ ($r=\infty$); this can be achieved by redefining $z_h \to a z_h$ as in~\eqref{eqn:zLinearSymm} (leaving $z_h e^{\psi_h}$ fixed), so $z_h$ is not a separate free parameter.

This counting implies a weak ``no hair'' theorem, as follows: for a fixed charge $Q_a$ and choice of vacuum $\phi^i_\infty = \phi^i(r = \infty) = \phi^i(z = 0)$,\footnote{Here and henceforward, the subscript $\infty$ refers to spatial infinity, regardless of the choice of coordinates.} one naively expects a unique black hole (black brane) solution for every mass (mass density) $\cM$ above the extremality bound. We found a family of solutions with $N+1$ free parameters $\phi_h^i$ and $z_h e^{\psi_h}$, which is the same number of parameters as $\phi^i_\infty$ and $\cM$. Indeed, for large $\cM$ (small $z_h e^{\psi_h}$) there is a one-to-one map between $(\phi_h^i, z_h e^{\psi_h})$ and $(\phi^i_\infty, \cM)$, so there is a unique solution for each mass.

This does not mean that our naive expectation is fulfilled. 
Indeed, as shown in~\cite{ExtPaper}, for some choices of the charge function $\cQ^2(\phi)$ there can be multiple (even infinitely many) solutions with the same mass. However, generally there are not continuous families of such solutions, in agreement with the above counting argument.

\bigskip

Now consider the case $r_h = 0$, so that $f=1$. The horizon is at $z_h = \infty$, and the metric function $R(\rho)$ in~\eqref{eqn:rhoYRFdef} must remain finite there for it to be smooth. Defining $\chi \df \psi + \log z$, we find $e^{-\chi} = (d-p-2) R^{d-p-2}$, so $\chi$ must remain finite at the horizon. Written in terms of $\chi$, the equations~\eqref{eqn:zEOM} become
\begin{align}
  \ddot{\chi} &=  \frac{1}{z^2} \biggl(\mathcal{Q}^2 (\phi) e^{2 \chi}-1\biggr) \,, & 
  \ddot{\phi}^i + \Gamma^i_{\; j k} (\phi)  \dot{\phi}^j  \dot{\phi}^k &=
  \frac{1}{2 z^2} \mathcal{G}^{i j} (\phi) \mathcal{Q}^2_{, j} (\phi) e^{2
  \chi}, \nonumber \\ & &
  \dot{\chi}^2-\frac{2}{z} \dot{\chi} +  \mathcal{G}_{i j}
  (\phi) \dot{\phi}^i  \dot{\phi}^j &=  \frac{1}{z^2} \biggl(\mathcal{Q}^2 (\phi) e^{2 \chi} - 1\biggr) .
\end{align}
In particular, because the double integral of $1/z^2$ is logarithmically divergent, integrating the first equation twice we conclude that a smooth horizon with finite $\chi_h = \chi(z = \infty)$ requires $\mathcal{Q}^2 (\phi) e^{2 \chi} \to 1$ as $z \to \infty$. Likewise, the second equation implies that $\mathcal{G}^{i j} (\phi) \mathcal{Q}^2_{, j} (\phi) e^{2 \chi} \to 0$ as $z \to \infty$ is required. Thus, the conditions
\begin{align} \label{eqn:attractorconds}
\cQ^2 (\phi_h) e^{2 \chi_h}&=1 \,, & \mathcal{Q}^2_{, j} (\phi_h) &= 0 \,,
\end{align}
are necessary for a smooth $r_h = 0$ horizon to exist. The second condition implies that the moduli reach a critical point of the charge function $\cQ^2(\phi)$ at the horizon, regardless of their asymptotic values, which is the well-known attractor mechanism~\cite{Ferrara:1995ih,Cvetic:1995bj,Strominger:1996kf,Ferrara:1996dd,Ferrara:1996um}. 

For solutions satisfying~\eqref{eqn:attractorconds}, $\chi$ approaches a finite value $\chi_h = - \log \cQ_h$ at the horizon where $\cQ_h^2 \df \cQ^2(\phi_h)$. Thus, the near-horizon geometry is $\mathrm{AdS}_{p + 1} \times S^{d - p - 1}$
\begin{equation}
  d s^2 \rightarrow R_{\mathrm{AdS}}^2  \left[ \frac{1}{w^2}  (- d t^2 + d w^2 +
  \delta_{m n} d y^m d y^n) + \frac{(d - p - 2)^2}{p^2} d \Omega^2_{d - p - 1}
  \right], \label{eqn:nearhorizon}
\end{equation}
where
\begin{equation}
  R_{\mathrm{AdS}} \df \frac{p e^{- \frac{1}{d - p - 2} \chi_h}}{(d - p -
  2)^{\frac{d - p - 1}{d - p - 2}}}, \qquad w \df R_{\mathrm{AdS}} e^{-
  \frac{1}{p} \chi_h } z^{\frac{1}{p}} .
\end{equation}
This includes all smooth $r_h = 0$ horizons, but horizons of this kind with $p>1$ are not smooth, as previously discussed.

To extend these solutions outside the near-horizon region, first consider the case $\phi^i(z) = \phi^i_h$. Then~\eqref{eqn:zCons} integrates to
\begin{equation}
\psi = - \log(\cQ_h (z + a))
\end{equation}
upon imposing the requirement that the solution is regular at large positive $z$, where $\cQ_h \df \cQ(\phi_h) > 0$ and $a$ is an integration constant to be determined. This is an extremal Reissner-Nordstr\"om solution in an unusual gauge. Other $r_h = 0$ solutions can be constructed from this one perturbatively in small $\delta \phi^i(z) = \phi^i(z) - \phi^i_h$. In particular, to first order~\eqref{eqn:phizEOM} gives
\begin{equation} \label{eqn:deltaphi}
\delta\ddot{\phi}^i = \frac{1}{\cQ_h (z+a)^2} \cG_h^{i j} \cQ^h_{jk} \delta \phi^k \,,
\end{equation}
where $\cQ^h_{i j} \df \cQ_{,ij}(\phi_h)$ and $\cG_h^{i j} \df \cG^{i j}(\phi_h)$.
We choose local coordinates on the scalar manifold such that $\cG^{i j}(\phi_h) = \delta^{i j}$ and $\cQ_{,i j}(\phi_h) = 0$ for $i \ne j$. This simplifies~\eqref{eqn:deltaphi} to $(z+a)^2 \delta\ddot{\phi}^i = \frac{\cQ^h_{i i}}{\cQ_h} \delta \phi$, with the general solution
\begin{equation} \label{eqn:deltaphisoln}
\delta \phi^i = A_i (z+a)^{\frac{1}{2} + \sqrt{\frac{1}{4}+ \frac{\cQ^h_{i i}}{\cQ_h}}} + B_i (z+a)^{\frac{1}{2} - \sqrt{\frac{1}{4}+ \frac{\cQ^h_{i i}}{\cQ_h}}} \,.
\end{equation}
Assuming $\cQ^h_{ii} > 0$, the first exponent is positive and the second is negative, so $\delta \phi^i \to 0$ as $z \to \infty$ requires $A_i = 0$. Imposing $\psi(z=0) = 0$ and $\delta\phi^i(z=0) = \delta\phi^i_\infty$, we obtain
\begin{subequations} \label{eqn:pertsoln}
\begin{align}
\psi(z) &= - \log(1+\cQ_h z)+O(\delta\phi_\infty^2)\,, \\ \phi^i(z) &= \phi^i_h + \delta\phi^i_\infty (1+\cQ_h z)^{\frac{1}{2} - \sqrt{\frac{1}{4}+ \frac{\cQ^h_{i i}}{\cQ_h}}}+O(\delta\phi_\infty^2) \,,
\end{align}
\end{subequations}
to linear order. If instead $\cQ^h_{ii} < 0$ for some $i$, then both terms in~\eqref{eqn:deltaphisoln} grow as $z \to \infty$. This implies $A_i = B_i = 0$, so that $\delta \phi^i = 0$ to this order. Thus, $\cQ^2(\phi)$ cannot increase as the horizon is approached, similar to~\eqref{eqn:smoothhorizon}.

The perturbative solution~\eqref{eqn:pertsoln} can be systematically improved order by order in $\delta \phi_\infty$. More generally, any $r_h = 0$ solution will take this form close enough to the horizon, where $\delta \phi \to 0$. 
A different approach to constructing $r_h=0$ solutions is discussed in~\cite{ExtPaper}.

\begin{figure}
\begin{center}
\includegraphics{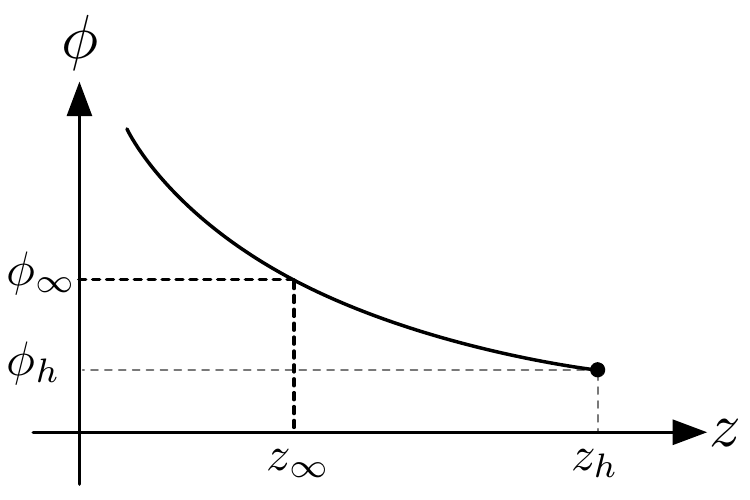}
\caption{Solutions to the equations~\eqref{eqn:zEOM} come in one-parameter families related by the symmetry~\eqref{eqn:zLinearSymm}. Given one solution $\phi(z)$ in the vacuum $\phi_\infty^{(0)} = \phi(0)$, pick any $z_\infty < z_h$ and set $z_\infty = 0$ using~\eqref{eqn:zLinearSymm}. This gives a new solution in the vacuum $\phi_\infty = \phi(z_\infty)$ with the same value of $\phi_h$. \label{fig:Zfamilies}}
\end{center}
\end{figure}

Note that for both $r_h > 0$ and $r_h = 0$ cases, solutions with the same moduli values $\phi_h^i$ at the horizon are related in one-parameter families by the symmetry $z \to a z + b$ in \eqref{eqn:zLinearSymm}, where $a = a(b)$ is chosen to set $\psi_\infty = \psi(z=0) = 0$. The net effect of this is to change the location of the asymptotic boundary $z=0$, thereby changing $\phi_\infty^i$ (tracking the profile $\phi^i(z)$) while holding $\phi_h^i$ fixed. This is illustrated in Figure~\ref{fig:Zfamilies}.

\subsection{Thermodynamics and long range forces} \label{subsec:BHthermo}

The ADM mass density of the solution can be computed using (2.8) of \cite{lu:1993vt}, giving:
\begin{equation}
  \mathcal{M}= \frac{V_{d - p - 1}}{\kappa_d^2}  \biggl[ - \frac{1}{\xi} 
  \dot{\psi}_{\infty} + \frac{d - p - 1}{d - p - 2} \cdot \frac{1}{2 z_h}
  \biggr] \,, \label{eqn:ADMmass}
\end{equation}
where $\dot{\psi}_{\infty} = \dot{\psi}(z=0)$. For $p = 1$, this is the ADM mass of the black hole. For $p > 1$, there is also an
ADM tension:
\begin{equation}
  \mathcal{T}= \frac{V_{d - p - 1}}{\kappa_d^2}  \left[ - \frac{1}{\xi} 
  \dot{\psi}_{\infty} + \frac{1}{d - p - 2} \cdot \frac{1}{2 z_h} \right],
  \label{eqn:ADMtension}
\end{equation}
see, e.g., \cite{Townsend:2001rg}. This can be derived by similar methods as (2.8) in
\cite{lu:1993vt} using the gravitational stress-energy ``tensor'' referenced to
a flat background and integrated over the transverse directions. Alternatively, we can compare the $r\to \infty$ limit of the solution with the linearized result~\eqref{eqn:farfield}. Note that
$\mathcal{T} \leqslant \mathcal{M}$ with equality if and only if $r_h = 0$.

The area of the black hole event horizon can be read off from the metric ansatz~\eqref{eqn:finalmetricansatz} with $p=1$,
\begin{equation}
A=r^{d-2}_h e^{-\frac{d-2}{d-3} \psi_h} V_{d-2} \,. \label{eqn:horizonarea}
\end{equation}
In the black brane case, this generalizes to the horizon area per unit $y$ volume along the brane
\begin{equation}
  \mathcal{A}= r^{d - p - 1}_h e^{- \psi_h / \xi} V_{d - p - 1} \,,
  \label{eqn:hArea}
\end{equation}
where $e^{- \psi_h / \xi}  = e^{\frac{p-1}{p} \psi_h - \frac{d-p-1}{d-p-2}\psi_h}$ accounts for the warping in both the $y$ and $S^{d-p-1}$ directions.

To compute the horizon surface gravity, we write
the metric in infalling coordinates 
\begin{equation}
  d s^2 = - f e^{\frac{2}{p} \psi} d v^2 + 2 e^{\left[ \frac{1}{p} -
  \frac{1}{d - p - 2} \right] \psi} d v d r + e^{\frac{2}{p} \psi} \delta_{i
  j} d y^i d y^j + e^{- \frac{2}{d - p - 2} \psi} r^2 d \Omega^2_{d - p - 1} .
\end{equation}
The surface gravity $\kappa$ is the solution to the equation
\begin{equation}
  k^a \nabla_a k^b = \kappa k^b, \qquad k = \frac{\partial}{\partial v},
\end{equation}
on the horizon. Since $k$ is a Killing vector, this is the same as $\nabla_b
k^2 = - 2 \kappa k_b$, or
\begin{equation}
  \kappa = \left. - \frac{1}{2 g_{v r}} g_{v v, r} \right|_{r = r_h} = \frac{d
  - p - 2}{2 r_h} e^{\psi_h / \xi} . \label{eqn:hGrav}
\end{equation}
Notice in particular that
\begin{equation} \label{eqn:kappaA}
\kappa \mathcal{A} = \frac{V_{d-p-1}}{2z_h} \,.
\end{equation}
Therefore, solutions with smooth horizons are quasiextremal if and only if $r_h = 0$ ($z_h = \infty$).

Comparing~\eqref{eqn:ADMmass} and~\eqref{eqn:ADMtension} with~\eqref{eqn:kappaA}, we find:
\begin{equation} \label{eqn:blackbraneconstitutive}
\mathcal{T} = \mathcal{M} - \frac{1}{\kappa_d^2} \kappa \mathcal{A} \,,
\end{equation}
where the surface gravity $\kappa$ should not be confused with the Einstein constant $\kappa_d^2$.
This is very similar to the perfect brane relation~\eqref{eqn:braneconstitutive}. In fact, the $d$-dimensional analogs of the Hawking temperature and the Bekenstein-Hawking entropy are (see, e.g.,~\cite{myers:1986un})
\begin{align} \label{eqn:BekensteinHawking}
T &= \frac{\kappa}{2\pi}\,, & S &= \frac{2 \pi}{\kappa_d^2} A \,,
\end{align}
so~\eqref{eqn:blackbraneconstitutive} exactly matches~\eqref{eqn:braneconstitutive}, where $s = \frac{2 \pi}{\kappa_d^2} \cA$ is the entropy density of the black brane.

From~\eqref{eqn:PhiInt0}, the electrostatic potential is
\begin{equation}
  \Phi^a (z) = \frac{1}{V_{d - p - 1}}  \int_0^z \f^{a b} (\phi) Q_b e^{2 \psi} d z \, . \label{eqn:ESpot}
\end{equation}
In particular,
\begin{equation}
\Phi_h^a Q_a = \frac{V_{d - p - 1}}{\xi \kappa_d^2}  \int_0^{z_h}
  \mathcal{Q}^2 (\phi) e^{2 \psi} d z = \frac{V_{d - p - 1}}{\xi \kappa_d^2} 
  \int_0^{z_h} d_z [f \dot{\psi}] d z = - \frac{V_{d - p - 1}}{\xi \kappa_d^2}
  \dot{\psi}_{\infty} \,. \label{eqn:PhiQ}
\end{equation}
For $r_h > 0$, we used the fact that $\dot{\psi}$ remains finite at the horizon (see~\eqref{eqn:smoothhorizon}). In the quasiextremal case we instead use $\dot{\psi} \to 0$ as $z \to \infty$ (because $\chi = \psi + \log z$ approaches a constant value).

Comparing~\eqref{eqn:PhiQ} and~\eqref{eqn:kappaA} with~\eqref{eqn:ADMmass}, we obtain the Smarr formula
\begin{equation} \label{eqn:braneSmarr}
  \mathcal{M} = \Phi_h^a Q_a + \frac{d - p - 1}{d - p - 2} \cdot
  \frac{1}{\kappa_d^2} \kappa \mathcal{A}\,.
\end{equation}
This reproduces the results of, e.g., \cite{Townsend:2001rg} (in the non-spinning case), here generalized to include arbitrary moduli.

The Smarr formula has a simple application to cosmic censorship, as follows. Suppose that we attempt to overcharge a quasiextremal black hole by dropping in a charged particle with charge $q_a = x Q_a$ for $x \ll 1$. By~\eqref{eqn:conservedE}, the particle must have energy $E \ge q_a \Phi_h^a$ to cross the event horizon, so by the Smarr formula $E \ge x Q_a \Phi_h^a = x M$ since either $\kappa = 0$ or $A =0$ (the Hawking temperature or the Bekenstein-Hawking entropy vanishes) for a quasiextremal black hole. Therefore, the charge-to-mass ratio of the resulting black hole cannot be larger than the initial, quasiextremal one.

To derive a ``first law of black hole mechanics'', we consider a solution to (\ref{eqn:zEOM}) with a smooth
horizon and perturb the solution infinitesimally to find a nearby solution.
By perturbing~\eqref{eqn:psizEOM} and (\ref{eqn:psiNoQ}), we find that the infinitesimal perturbation satisfies the linear ODEs:
\begin{subequations}
\begin{align}
  d_z [f \delta \dot{\psi} + \delta\! f \dot{\psi}] &=  \mathcal{Q}^2_{, i}
  (\phi) \delta \phi^i e^{2 \psi}+ 2 \delta \psi \mathcal{Q}^2 (\phi) e^{2 \psi} + 2
  \xi \frac{\kappa_d^2 f^{a b} (\phi) Q_a \delta Q_b}{V_{d - p - 1}^2} e^{2
  \psi} \,, \label{eqn:linearEOM1} \\  
  \delta \ddot{\psi} &= 2 \dot{\psi} \delta \dot{\psi} +  \mathcal{G}_{i
  j, k} (\phi) \dot{\phi}^i  \dot{\phi}^j \delta \phi^k + 2  \mathcal{G}_{i
  j} (\phi) \dot{\phi}^i  \delta \dot{\phi}^j  \,, \label{eqn:linearEOM2}
\end{align}
\end{subequations}
where $\delta\! f = \frac{z}{z_h^2} \delta z_h = (1 - f)  \frac{\delta
z_h}{z_h}$.
Adding~\eqref{eqn:linearEOM1} to $f$ times~\eqref{eqn:linearEOM2}
and simplifying using the background equations of motion, we obtain:
\begin{equation}
  d_z [2 f \delta \dot{\psi} + \delta\! f \dot{\psi}] - \dot{f}  \delta \dot{\psi} = 2 d_z [f \dot{\psi} \delta \psi] + 2 d_z [\mathcal{G}_{i j}
  (\phi) f \dot{\phi}^i \delta \phi^j] + 2 \xi \frac{\kappa_d^2 f^{a b}
  (\phi) Q_a \delta Q_b}{V_{d - p - 1}^2} e^{2 \psi} . \label{eqn:pertCombo}
\end{equation}
Using (\ref{eqn:ESpot}) and noting that $\ddot{f} = 0$, this can be integrated to give
\begin{equation}
  f \dot{\delta \psi} + \frac{1}{2} \delta\! f \dot{\psi} = \left( f \dot{\psi}
  + \frac{1}{2}  \dot{f} \right) \delta \psi + \mathcal{G}_{i j} (\phi) f
  \dot{\phi}^i \delta \phi^j + \frac{\xi \kappa_d^2}{V_{d - p - 1}}  (\Phi^a -
  \Phi_h^a) \delta Q_a + \frac{1}{2 z_h} \delta \psi_h \label{eqn:linearCons}
\end{equation}
where
the integration constant is fixed by evaluating the equation at the horizon $z =
z_h$ and using the identity
\begin{equation}
  \delta \psi_h = \delta \psi (z_h) + \dot{\psi} (z_h) \delta z_h .
\end{equation}
Evaluating $\left( \ref{eqn:linearCons} \right)$ at $z = 0$ ($r = \infty$), we
obtain
\begin{equation}
  \delta \dot{\psi}_{\infty} = \mathcal{G}_{i j}^{\infty} (\phi)
  \dot{\phi}^i_{\infty} \delta \phi^j_{\infty} - \frac{\xi \kappa_d^2}{V_{d -
  p - 1}} \Phi_h^a \delta Q_a + \frac{1}{2 z_h} \delta \psi_h \;,
\end{equation}
where we used $\psi_{\infty} = 0$.

To interpret this condition, note that per~\eqref{eqn:scalarchargedef}, the scalar charge is
\begin{equation}
  \mu_i \df - V_{d - p - 1} G_{i j}^{\infty} \dot{\phi}_{\infty}^j =  - \frac{V_{d - p - 1}}{\xi \kappa_d^2} \mathcal{G}^{\infty}_{i j} 
  \dot{\phi}_{\infty}^j. \label{eqn:scalarcharge1}
\end{equation}
Applying (\ref{eqn:ADMmass}), (\ref{eqn:hArea}), and (\ref{eqn:hGrav}), we obtain:
\begin{equation}
  \delta \mathcal{M}= \mu_i \delta \phi^i_{\infty} + \Phi_h^a \delta Q_a +
  \frac{1}{\kappa_d^2} \kappa \delta \mathcal{A}. \label{eqn:firstlaw}
\end{equation}
This is the first law of black hole / black brane mechanics in the non-spinning case, generalized to include scalars as in, e.g.,
\cite{Gibbons:1996af}. Written in terms of the temperature $T$ and entropy density $s$ using~\eqref{eqn:BekensteinHawking}, the last term takes the expected form $T \delta s$.

Using the first law, we obtain an alternate interpretation of the scalar charge
\begin{equation} \label{eqn:scalarcharge2}
\mu_i = \frac{\partial \cM}{\partial \phi^i_{\infty}}\biggr|_{\cA, Q_a, \phi_\infty^{j\ne i}} \,,
\end{equation}
in agreement with the perfect brane relation~\eqref{eqn:blackbranescalarcharge}.

Finally, we consider the long range force between two identical black holes / parallel black branes. In particular, \eqref{eqn:zCons} evaluated at $z=0$ gives
\begin{equation}
  \dot{\psi}_{\infty} \biggl(\dot{\psi}_{\infty} - \frac{1}{z_h}\biggr) + \frac{\xi \kappa_d^2}{V_{d-p-1}^2} G^{i j}_{\infty} \mu_i \mu_j =  \frac{\xi \kappa_d^2}{V_{d-p-1}^2} \f^{a b}_{\infty} Q_a Q_b \,,
\end{equation}
using~\eqref{eqn:scalarcharge1}. By comparison, using~\eqref{eqn:ADMmass} and~\eqref{eqn:ADMtension}, we obtain
\begin{multline}
\cT^{\mu \nu} \cT_{\mu \nu} - \frac{1}{d-2} \cT_{\; \mu}^{\mu} \cT_{\; \nu}^{\nu} = \cM^2 + (p-1) \cT^2 - \frac{(\cM+(p-1)\cT)^2}{d-2}  \\
= \frac{V_{d-p-1}^2}{\kappa_d^4} \Biggl[\frac{1}{\xi} \dot{\psi}_{\infty} \biggl(\dot{\psi}_{\infty} - \frac{1}{z_h}\biggr) + \frac{d-p-1}{d-p-2}\cdot \frac{1}{4 z_h^2} \Biggr] \,.
\end{multline}
Combining these two equations,
\begin{align}
\f^{a b}_{\infty} Q_a Q_b - G^{i j}_{\infty} \mu_i \mu_i  -\kappa_d^2 \biggl[\cT^{\mu \nu} \cT_{\mu \nu} - \frac{1}{d-2} \cT_{\; \mu}^{\mu} \cT_{\; \nu}^{\nu}\biggr] = - \frac{d-p-1}{d-p-2}\cdot \frac{V_{d-p-1}^2}{4 \kappa_d^2 z_h^2} \,.
\end{align}
Thus, by~\eqref{eqn:branepressure}, the long-range pressure (force) between the identical parallel branes (identical black holes) separated by a large distance $r$ is
\begin{align}
P = - (d-p-1)(d-p-2) \frac{V_{d-p-1}}{4 \kappa_d^2} \frac{r_h^{2(d-p-2)}}{r^{d-p-1}} \,.
\end{align}
Since $r_h \ge 0$ vanishes if and only if the solution is quasiextremal, we conclude that quasiextremal black holes and black branes have vanishing long-range self-force,\footnote{Note that the long-range force between identical quasiextremal black holes must vanish if a corresponding family of static multicenter solutions exists. Such solutions can be found quite generally using known methods, e.g.,~\cite{Breitenlohner:1987dg,Galtsov:1998mhf,Bergshoeff:2008be,VanRiet:2020csu}. I thank Thomas Van Riet for discussions on this point.} whereas non-extremal ones have an attractive long-range self-force. This generalizes a well-known property of Reissner-Nordstr\"om black holes to static spherically symmetric black holes and black branes in a large class of two-derivative theories with moduli.

The principal assumptions underlying this result are (1) that the deep infrared is described by a weakly coupled two-derivative effective action (2) with the general form~\eqref{eqn:generalaction} (motivated in~\secref{sec:setup}) upon truncating to the neutral bosons, as well as (3) that solutions do not enter a strongly coupled region of moduli space outside the event horizon.\footnote{I am not aware of any solutions of this type. It is unclear whether they exist, see footnote~\ref{fn:stronglycoupledBHs}.}
 (Although we ignored solutions that cross between different branches of the moduli space, these are addressed in~\cite{CornersPaper}.)

\section*{Acknowledgements}

I thank Mirjam Cveti\v{c}, Finn Larsen, Matthew Reece and Tom Rudelius for helpful discussions and Mirjam Cveti\v{c}, Muldrow Etheredge, Matteo Lotito, Miguel Montero, Matthew Reece, Tom Rudelius and Irene Valenzuela for comments on the manuscript. I am especially indebted to Matthew Reece and Tom Rudelius, who collaborated on the early stages of this project. This research was supported by National Science Foundation grant PHY-1914934 during its final stages, and by Perimeter Institute for Theoretical Physics during its inception. Research at Perimeter Institute is supported by the Government of Canada through the Department of Innovation, Science and Economic Development, and by the Province of Ontario through the Ministry of Research, Innovation and Science. I thank the Yau Mathematical Sciences Center and the Kavli Institute for the Physics and Mathematics of the Universe for hospitality during the later stages of this work. This work was performed in part at Aspen Center for Physics, which is supported by National Science Foundation grant PHY-1607611.

\appendix

\settocdepth{section}

\section{Magnetic charges, theta angles, and self-dual gauge fields} \label{app:democratic}

In dimension $D = 2p+2$, $(p-1)$-branes can carry both electric and magnetic charges. In the main text, I assumed that the magnetic charge vanished, which also decoupled the theta term. I now generalize the discussion to include magnetic charge and theta angles, as well as the possibility (for  $D=4k+2$) of self-dual gauge fields. To do so, it is convenient to formulate the gauge boson action democratically, with separate electric and magnetic potentials for each gauge field related by a constraint. Below, I first review the democratic formulation, then return to the question of long range forces.

\subsection{Democratic formulation}
 The most general two-derivative pseudo-action for gauge fields $F^a_{D/2} = d A_{D/2-1}^a$ takes the form:
\begin{equation}
  S = - \frac{1}{8 \pi} \int t_{a b} (\phi) F^a_{D / 2} \wedge \ast F^b_{D /
  2} - \frac{1}{8 \pi} \int \tilde{t}_{a b} (\phi) F^a_{D / 2} \wedge F^b_{D /
  2} , \label{eqn:pS0}
\end{equation}
where $t_{a b} (\phi)$ and $\tilde{t}_{a b}(\phi)$ are functions of the moduli.
Note that $t_{a b} = t_{b a}$, but $\tilde{t}_{a b} = \pm \tilde{t}_{b a}$,
where the upper (lower) sign corresponds to $D = 4 k$ ($D = 4 k + 2$).

The Euler-Lagrange equations associated with~\eqref{eqn:pS0} are:
\begin{equation}
  d [t_{a b} \ast F^b + \tilde{t}_{a b} F^b] = 0 .
\end{equation}
The electromagnetic duality constraints should be consistent with and  imply these
equations, 
so they should take the form
\begin{equation}
  t_{a b} \ast F^b + \tilde{t}_{a b} F^b = \eta_{a b} F^b,
\end{equation}
for some constant matrix $\eta_{a b}$. This can be rewritten as
  $\ast F^a = \Lambda^a_{\;\; b} F^b$,
where $\Lambda^a_{\;\;b} \df t^{a b}  (\eta_{a b} -
\tilde{t}_{ab})$ must satisfy $\Lambda^2 = \mp 1$ for consistency,
since $\ast^2 \omega_{D / 2} = \mp \omega_{D / 2}$.

The stress tensor is
\begin{equation}
  T_{m n} = \frac{1}{4 \pi} t_{a b}  (F^a \circ F^b)_{m n},
\end{equation}
where $\circ$ is defined in~\eqref{eqn:circdef} and
satisfies $\ast \omega \circ \ast \chi = \omega \circ \chi$. Therefore
\begin{align}
  T_{m n} &= \frac{1}{8 \pi} t_{a b}  [(F^a \circ F^b)_{m n} + (\ast F^a \circ
  \ast F^b)_{m n}] = \frac{1}{8 \pi} t_{a b}  [(F^a \circ F^b)_{m n} +
  ((\Lambda F)^a \circ (\Lambda F)^b)_{m n}] \nonumber \\ 
  &= \frac{1}{4 \pi}  \hat{t}_{a b}  (F^a \circ F^b)_{m n},  
\end{align}
where
  $\hat{t} \df \frac{1}{2}  (t + \Lambda^{\top} t \Lambda)$.
Note that
\begin{equation}
  \hat{t} \Lambda = \frac{1}{2}  (t \Lambda \mp \Lambda^{\top} t) =
  \hat{\eta}, \qquad \text{where} \qquad \hat{\eta} \df \frac{\eta \mp
  \eta^{\top}}{2},
\end{equation}
so that $\hat{\eta}^{\top} = \mp \hat{\eta}$. Thus, the stress-energy tensor,
self-duality constraint, and $F^a$ equations of motion (which follow from the
self-duality constraint) are equivalent to those derived from the
psuedo-action:
\begin{equation}
  \hat{S} = - \frac{1}{8 \pi} \int \hat{t}_{a b} (\phi) F^a \wedge \ast F^b,
  \qquad \text{with the constraint} \qquad \hat{t}_{a b} \ast F^b =
  \hat{\eta}_{a b} F^b,
\end{equation}
where $\hat{\eta}^{\top} = \mp \hat{\eta}$ is a constant matrix and
$\hat{t}_{a b} (\phi)$ is constrained to satisfy $(\hat{\eta}^{- 1} 
\hat{t})^2 = \mp 1$ (equivalent to $\Lambda^2 = \mp 1$).

To complete the picture, we check that the $\phi$ equations of motion are
likewise unchanged. Varying with respect to the scalars, we obtain
\begin{equation}
  \delta \Lambda = \delta t^{- 1} (\eta - \tilde{t}) - t^{- 1} \delta
  \tilde{t} = - t^{- 1} (\delta t \Lambda + \delta \tilde{t}),
\end{equation}
and so
\begin{equation}
  \delta \hat{t} = \frac{1}{2}  (\delta t + \Lambda^{\top} \delta t \Lambda) +
  \frac{1}{2}  (\Lambda^{\top} t \delta \Lambda + \delta \Lambda^{\top} t
  \Lambda) = \frac{1}{2}  (\delta t - \Lambda^{\top} \delta t \Lambda) -
  \frac{1}{2}  [\Lambda^{\top} \delta \tilde{t} + \delta \tilde{t}^{\top}
  \Lambda].
\end{equation}
Thus,
\begin{align}
  \delta \hat{t}_{a b} F^a \wedge \ast F^b & = \frac{1}{2} \delta t_{a b} 
  (F^a \wedge \ast F^b - \ast F^a \wedge \ast^2 F^b) - \frac{1}{2} \delta
  \tilde{t}_{a b} \ast F^a \wedge \ast F^b - \frac{1}{2} \delta \tilde{t}_{b
  a} F^a \wedge \ast^2 F^b \nonumber \\
  & = \delta t_{a b} F^a \wedge \ast F^b + \delta \tilde{t}_{a b} F^a \wedge
  F^b, 
\end{align}
and so
\begin{equation}
  \delta \hat{S} = - \frac{1}{8 \pi} \int \delta \hat{t}_{a b} F^a \wedge \ast
  F^b = - \frac{1}{8 \pi} \int t_{a b} F^a \wedge \ast F^b - \frac{1}{8 \pi}
  \int \tilde{t}_{a b} F^a \wedge F^b = \delta S,
\end{equation}
as required.

Therefore, without loss of generality we can use the simplified pseudo-action
\begin{equation}
  S = - \frac{1}{8 \pi} \int t_{a b} (\phi) F^a \wedge \ast F^b, \qquad
  \text{with the constraint} \qquad t_{a b} \ast F^b = \eta_{a b} F^b,
\end{equation}
where $\eta_{a b} = (- 1)^{\frac{D - 2}{2}} \eta_{b a}$ is a constant matrix
and $t_{a b} (\phi) = t_{b a} (\phi)$ is constrained to satisfy $(\eta^{- 1}
t)^2 = (- 1)^{\frac{D - 2}{2}}$.

\subsection{Dimensional reduction and quantization} \label{sec:demReduction}

To understand the quantum dynamics of this theory, we dimensionally reduce on
a circle of radius $R$ to $d= D-1$ dimensions, via the ansatz:
\begin{equation}
  F^a_{(D)} = \tilde{F}^a + \frac{1}{2 \pi R} G^a \wedge (d y + R B), \qquad
  A^a_{(D)} = A^a + \frac{1}{2 \pi R} C^a \wedge (d y + R B),
\end{equation}
where $B$ is the graviphoton with field strength $H=dB$, $\tilde{F}^a = d A^a \pm \frac{1}{2 \pi} C^a \wedge H$, $G^a = d C^a$, and the metric takes the form:
\begin{equation}
  d s_D^2 = e^{\frac{\sigma}{d - 2}} d s_d^2 + e^{- \sigma}  (d y + R B)^2 .
\end{equation}
Note that
  $d \tilde{F}^a = \pm \frac{1}{2 \pi} G^a \wedge H$.
We find:
\begin{equation}
  \ast_D F^a_{(D)} = e^{- \frac{d - 1}{2 (d - 2)} \sigma} \ast \tilde{F}^a
  \wedge (d y + R B) \pm \frac{1}{2 \pi R} e^{\frac{d - 1}{2 (d - 2)} \sigma}
  \ast G^a,
\end{equation}
hence the pseudo-action reduces to
\begin{equation}
  S = - \frac{R}{4} \int e^{- \frac{d - 1}{2 (d - 2)} \sigma} t_{a b} 
  \tilde{F}^a \wedge \ast \tilde{F}^b - \frac{1}{16 \pi^2 R}  \int e^{\frac{d
  - 1}{2 (d - 2)} \sigma} t_{a b} G^a \wedge \ast G^b,
\end{equation}
with the constraints
\begin{equation}
  \pm \frac{1}{2 \pi R} e^{\frac{d - 1}{2 (d - 2)} \sigma} t_{a b} \ast G^b =
  \eta_{a b}  \tilde{F}^b, \qquad e^{- \frac{d - 1}{2 (d - 2)} \sigma} t_{a b}
  \ast \tilde{F}^b = \frac{1}{2 \pi R} \eta_{a b} G^b .
\end{equation}
It is easy to check that these two constraints are equivalent. Likewise, one
can check that the constraints imply the equations of motion. We now add a
total derivative to the action:
\begin{equation}
  S = - \frac{R}{4} \int e^{- \frac{d - 1}{2 (d - 2)} \sigma} t_{a b} 
  \tilde{F}^a \wedge \ast \tilde{F}^b - \frac{1}{16 \pi^2 R}  \int e^{\frac{d
  - 1}{2 (d - 2)} \sigma} t_{a b} G^a \wedge \ast G^b + \frac{1}{4 \pi} \int
  \eta_{a b} F^a \wedge G^b .
\end{equation}
This has no effect on the equations of motion. However, varying with respect
to $F^a = d A^a$, we obtain:
\begin{equation}
  \delta S = \frac{R}{2} \int \delta F^a \wedge \left[ - e^{- \frac{d - 1}{2
  (d - 2)} \sigma} t_{a b} \ast \tilde{F}^b + \frac{1}{2 \pi R} \eta_{a b} G^b
  \right],
\end{equation}
so if we treat $F^a$ as a fundamental field it acts as a Lagrange multiplier
imposing the constraint. Moreover, the missing Bianchi identity $d \tilde{F}^a
= \pm \frac{1}{2 \pi} G^a \wedge H$ follows from the constraint and the $G^a$
equation of motion. In this way, we recover a genuine action.

Integrating out the now-auxilliary $F^a$, we finally obtain:
\begin{equation}
  S = - \frac{1}{8 \pi^2 R} \int e^{\frac{d - 1}{2 (d - 2)} \sigma} t_{a b}
  G^a \wedge \ast G^b \mp \frac{1}{8 \pi^2} \int \eta_{a b} C^a \wedge G^b
  \wedge H . \label{eqn:genuineRed}
\end{equation}
Dirac quantization for this action gives the conditions $ \frac{1}{2 \pi} \oint G^a \in \mathbb{Z}$ and:
\begin{equation}
   \pm\frac{1}{4 \pi^2 R}  \oint \left[ e^{\frac{d - 1}{2 (d - 2)} \sigma} t_{a b}
  \ast G^b - \eta_{a b} R B \wedge G^b \right] = \frac{1}{2 \pi} \eta_{a b} 
  \oint \left[  \tilde{F}^b + \frac{1}{2 \pi} G^b \wedge B \right] \in
  \mathbb{Z}.
\end{equation}
Lifting back to $D$ dimensions, these conditions correspond to
\begin{equation}
  \frac{1}{2 \pi} \oint_{\alpha} F^a_{(D)} \in \mathbb{Z} \qquad \text{and}
  \qquad \eta_{a b}  \frac{1}{2 \pi} \oint_{\beta} F^b_{(D)} \in \mathbb{Z},
\end{equation}
for cycles $\alpha$ and $\beta$ respectively wrapping and not wrapping the
compact circle. For a genuine theory, the quantization condition must be the
same for both kinds of cycles, hence $\eta_{a b}$ must be a unimodular
matrix.\footnote{Invariance of the Chern-Simons term $\mp \frac{1}{8 \pi^2}
\int \eta_{a b} C^a \wedge G^b \wedge H$ under large gauge transformations
seems to suggest that $\eta_{a b}$ should be \emph{even}, but this can be cancelled by the Green-Schwarz mechanism, so the constraint is not universal. For instance, it is violated in 10d type IIB string
theory.}

In $D = 4 k$, $\eta_{a b}$ is unimodular and antisymmetric, from which it
follows that it can be put into the canonical form $\eta =
\left(\begin{array}{cc}
  0_{n \times n} & 1_{n \times n}\\
  - 1_{n \times n} & 0_{n \times n}
\end{array}\right)$ by a Gram-Schmidt-like process.\footnote{Start with a
primitive lattice vector $v^1$. Pick another vector $w_1$ such that $\langle
v^1, w_1 \rangle = 1$. Now pick $V^2$ which is LI from $v^1, w_1$, and define
$v^2 \df V^2 - w_1  \langle v^1, V^2 \rangle - \langle V^2, w_1 \rangle
v^1$ (if $v_2$ is not primitive, choose the primitive vector in this
direction). Next, pick $W_2$ such that $\langle v^2, W_2 \rangle = 1$, and
define $w_2 = W_2 - w_1  \langle v^1, W^2 \rangle - \langle W^2, w_1 \rangle
v^1$. Proceeding in this fashion, we obtain the desired basis.}
The duality group preserving this form is $\Sp{2 n, \mathbb{Z}}$.

In $D = 4 k + 2$, since $\eta_{a b}$ is unimodular and symmetric, it defines a
unimodular lattice, which is either even or odd. The signature of this lattice
specifies the number of self-dual and anti-self-dual chiral bosons. Likewise,
the automorphism group of this lattice is the duality group, which is a finite
group in the purely self-dual or purely anti-self dual cases. In the mixed
signature cases, there are only two possible lattices: the odd lattice $I_{m,
n}$ and the even lattice $\II_{m, n}$, where the latter only exists in
signature $m - n \equiv 0 \pmod  8$. The corresponding duality
groups can be denoted
  $\gO{m, n ; \mathbb{Z}}_I$ and  $\gO{m, n ; \mathbb{Z}}_{\II}$.
Finally, note that the symmetric matrices
\begin{equation}
  \eta_+ \df \frac{1}{2}  (t + \eta), \qquad \eta_- \df \frac{1}{2} 
  (t - \eta),
\end{equation}
are both positive semidefinite, since they result from applying projections
$\Pi_{\pm} \df \frac{1}{2}  (1 \pm \Lambda)$ to $t$, which is positive
definite. We have
\begin{equation}
  \eta = \eta_+ - \eta_-, \qquad t = \eta_+ + \eta_-,
\end{equation}
so the lattice data is equivalently represented in terms of
 ``left-moving'' and ``right-moving'' parts, $\eta_+$ and $\eta_-$ respectively, as is familiar on the string worldsheet. 

\subsection{Relation to non-democratic formulations}

Suppose that
\begin{equation}
  \eta = \left(\begin{array}{cc}
    0_{n \times n} & 1_{n \times n}\\
    \mp 1_{n \times n} & 0_{n \times n}
  \end{array}\right) .
\end{equation}
As explained above, this is always true in some basis in the case $D = 4 k$,
whereas it is possible in $D = 4 k + 2$ when $\eta$ is even\footnote{Such a basis also exists when $\eta$ is odd, but with non-integral quantization. This can be interpreted as a discrete theta angle in the resulting non-democratic action.} with signature
$(n, n)$. In this basis, the constraint $(\eta^{- 1} t)^2 = \mp 1$ has the
general solution:
\begin{equation}
  t = \begin{pmatrix}
    \tau_2 + \tau_1^{\top} \tau_2^{- 1} \tau_1 & - \tau_1^{\top} \tau_2^{-
    1}\\
    - \tau_2^{- 1} \tau_1 & \tau_2^{- 1}
  \end{pmatrix}, \label{eqn:ttotau}
\end{equation}
where $\tau_1^{\top} = \pm \tau_1$ and $\tau_2^{\top} = \tau_2$. The
pseudo-action is then:
\begin{equation}
  S = - \frac{1}{8 \pi} \int [\tau_2 + \tau_1^{\top} \tau_2^{- 1}
  \tau_1]_{\alpha \beta} F^{\alpha} \wedge \ast F^{\beta} + \frac{1}{4 \pi}
  \int [\tau_2^{- 1} \tau_1]_{\; \beta}^{\alpha} G_{\alpha} \wedge \ast
  F^{\beta} - \frac{1}{8 \pi} \int [\tau_2^{- 1}]^{\alpha \beta} G_{\alpha}
  \wedge \ast G_{\beta},
\end{equation}
with the constraints
\begin{equation}
  [\tau_2 + \tau_1^{\top} \tau_2^{- 1} \tau_1]_{\alpha \beta} \ast F^{\beta} -
  [\tau_1^{\top} \tau_2^{- 1}]_{\alpha}^{\;  \; \beta} \ast G_{\beta} =
  G_{\alpha}, \qquad - [\tau_2^{- 1} \tau_1]^{\alpha}_{\; \beta} \ast
  F^{\beta} + [\tau_2^{- 1}]^{\alpha \beta} \ast G_{\beta} = \mp F^{\alpha},
\end{equation}
which simplify to
  $G = \tau_1 F + \tau_2 \ast F$.
As in~\secref{sec:demReduction}, we add a total derivative $\mp \frac{1}{4 \pi} \int G_{\alpha} \wedge
F^{\alpha}$ to the pseduo-action and then observe that varying with respect to $G_{\alpha}$ gives
\begin{equation}
  \delta S = \frac{1}{4 \pi} \int [\tau_2^{- 1}]^{\alpha \beta} \delta
  G_{\alpha} \wedge \ast (\tau_1 F + \tau_2 \ast F - G)_{\beta},
\end{equation}
which is the constraint. Thus, we recover a genuine action by treating
$G_{\alpha}$ as a fundamental (auxilliary) field. Integrating it out, we are
left with:
\begin{equation}
  S = - \frac{1}{4 \pi} \int [\tau_2]_{\alpha \beta} F^{\alpha} \wedge \ast
  F^{\beta} - \frac{1}{4 \pi} \int [\tau_1]_{\alpha \beta} F^{\alpha} \wedge
  F^{\beta},
\end{equation}
which is the usual action with gauge kinetic matrix $\f_{\alpha \beta} =
\frac{1}{2 \pi}  (\tau_2)_{\alpha \beta}$ and theta angle $\theta_{\alpha
\beta} = 2 \pi [\tau_1]_{\alpha \beta}$. The same steps in reverse democratize a standard Maxwell action.

In $D = 4 k + 2$ with $\eta$ of signature $(p, q)$ for $p \neq q$, it is not
posssible to write down a standard Maxwell action. Of course, we can describe
up to $\min (p, q)$ bosons in terms of unconstrained gauge fields, leaving at
minimum $| p - q |$ constraints. There is not much benefit to this hybrid
approach, but it is worth noting that we can easily go back and forth in the
same manner as above. In particular, we can democratize a hybrid action as
follows. First, ignoring the constraints we democratize the underlying
pseudo-action. The original constraints can now be expressed in terms of the
democratic $F^a$s with no $\ast F^a$s, and so can be used to algebraically
eliminate a like number of gauge fields from the new psuedo-action, resulting
in a fully democratic action.

\subsection{Long range forces}

With a thorough understanding of the democratic approach in hand, I now turn to the problem of long range forces between black holes and black branes.

We couple a probe Dirac brane to the above democratic action via the usual
action:
\begin{equation}
  S_{\text{brane}} = - \int d^{\frac{D - 2}{2}} \xi \sqrt{- \tilde{g}}
  \mathcal{T} (\phi) - Q_a \int A^a . \label{eqn:probebrane}
\end{equation}
This action is quantum mechanically sensible for $Q_a \in \mathbb{Z}$ because
$e^{i S}$ is invariant under large gauge transformations of the background
gauge field $A^a$.

However, attempting to derive the backreaction of the brane on the background
fields from this action leads to an immediate puzzle: the action
(\ref{eqn:probebrane}) implies that the branes are electrically charged but
not magnetically charged, i.e.,
\begin{equation}
  \frac{1}{2 \pi} d [t_{a b} (\phi) \ast F^b] = \mp 2 Q_a j_{\text{brane}},
  \qquad \frac{1}{2 \pi} d F^a = 0, \label{eqn:naivebackreaction}
\end{equation}
where the \emph{brane current} $j_{\text{brane}}$ is a delta-function
supported form with the defining property:
\begin{equation}
  \int_{\text{brane}} \omega_p = \int_{\text{spacetime}} \omega_p \wedge
  j_{\text{brane}} .
\end{equation}
(\ref{eqn:naivebackreaction}) is obviously incompatible with the
constraints $t_{a b} \ast F^b = \eta_{a b} F^b$. To correct it, we manually
symmetrize over the self-duality constraint to obtain
\begin{equation}
  \frac{1}{2 \pi} d [t_{a b} (\phi) \ast F^b] = \mp Q_a j_{\text{brane}},
  \qquad \frac{1}{2 \pi} \eta_{a b} d F^b = \mp Q_a j_{\text{brane}},
  \label{eqn:yesEOM}
\end{equation}
where the extra factor of $1 / 2$ in the first equation is notable (the other
half of the charge being magnetic).

To justify (\ref{eqn:yesEOM}), we begin with a heuristic argument, followed
later by more precise arguments. Note that (\ref{eqn:yesEOM}) follows from the
pseudo-action:
\begin{equation}
  S = - \frac{1}{8 \pi} \int t_{a b} (\phi) F^a \wedge \ast F^b - \int
  d^{\frac{D - 2}{2}} \xi \sqrt{- \tilde{g}} \mathcal{T} (\phi) - \frac{1}{2}
  Q_a \int A^a,
\end{equation}
with the modified Bianchi identity $d F^a = \mp 2 \pi [\eta^{- 1} Q]^a
j_{\text{brane}}$. Despite the $1 / 2$ in the last term, this action produces
the same brane dynamics as before, because the Maxwell term also depends on
the brane position:
\begin{multline}
  \delta S_{\text{Max}} = - \frac{1}{4 \pi} \int t_{a b} (\phi) \delta F^a
  \wedge \ast F^b = \frac{1}{4 \pi} \int \eta_{a b} F^a \wedge \delta F^b =
  \pm \frac{1}{4 \pi} \int \eta_{a b} A^a \wedge d (\delta F^b) \\ = -
  \frac{1}{2} Q_a \int A^a \wedge \delta j_{\text{brane}},
\end{multline}
where we apply the constraint in the second step and use $\delta (d F^a) = \mp
2 \pi [\eta^{- 1} Q]^a \delta j_{\text{brane}}$ in the last step. This extra
contribution doubles the effect of the term $\frac{1}{2} Q_a \int A^a$, and
reproduces the same dynamics as the probe action.\footnote{Note that this argument is still heuristic, as I am glossing over
some important subtleties.}

To further justify (\ref{eqn:yesEOM}), we appeal to non-democratic
formulations, in one of two ways: (1) if the Dirac self-pairing $Q_a  [\eta^{-
1}]^{a b} Q_b$ vanishes (always the case in $d = 4 k$), then we can choose a
description (undemocratic or partially democratic) where the gauge field $Q_a
A^a$ is unconstrained, and the above simply follow from backreacting the probe
action directly. Alternatively, even when $Q_a  [\eta^{- 1}]^{a b} Q_b \neq 0$,
we can reduce on a circle along the brane world-volume, producing the new
probe brane action:
\begin{equation}
  S_{\text{brane}} = (\ldots) - Q_a \int C^a . \label{eqn:probebrane2}
\end{equation}
Coupled to (\ref{eqn:genuineRed}), we find:
\begin{equation}
  \frac{1}{4 \pi^2 R} d \left[ e^{\frac{d - 1}{2 (d - 2)} \sigma} t_{a b} \ast
  G^b \right] - \frac{1}{4 \pi^2} \eta_{a b} G^b \wedge H = \pm Q_a
  j_{\text{brane}}^{(d)},
\end{equation}
or
\begin{equation}
  \frac{1}{2 \pi} \eta_{a b} d \biggl[ \tilde{F}^b + \frac{1}{2 \pi} G^b \wedge
  B \biggr] = Q_a j_{\text{brane}}^{(d)} .
\end{equation}
In combination with $d G^a = 0$, this lifts to
\begin{equation}
  \frac{1}{2 \pi} \eta_{a b} d F^b = \mp Q_a j_{\text{brane}}^{(D)},
\end{equation}
where it should be noted that $j_{\text{brane}}^{(d)} = \mp
j_{\text{brane}}^{(D)}$ in the case of a brane that wraps the compact circle,
the sign relating to the definition of positive orientation in each
case.\footnote{We take $\int^{(p)} \omega_p > 0$ when $\int^{(p + 1)} \omega_p
\wedge d y > 0$ for both the brane worldvolume and spacetime orientations.
Thus, for branes wrapping the circle we get a $\mp$ sign from $d y \wedge
j_{\text{brane}} = \mp j_{\text{brane}} \wedge d y$.} The self-duality
constraint then implies,
\begin{equation}
  \frac{1}{2 \pi} d [t_{a b} (\phi) \ast F^b] = \mp Q_a
  j_{\text{brane}}^{(D)},
\end{equation}
as we guessed before, which is half of the naively expected result. This
factor of $1 / 2$ is easily checked against the well-understood case of D3
branes in type IIB string theory.

With this caveat in mind, we can now calculate the long-range force between
parallel branes. 
The calculation is completely analogous to that of~\secref{sec:braneselfforce} upon treating the
pseudo-action as a genuine action with $f_{a b} = \frac{1}{4 \pi} t_{a b}$,
except that the electric backreaction of the brane includes an additional
factor of $1 / 2$. (Although the brane now backreacts magnetically as well,
this has no effect on the force between stationary branes.) We obtain
\begin{equation}
  \mathcal{P}_{12} = 2 \pi t^{a b} Q_{1a} Q_{2b} - G^{i j} \partial_i \mathcal{T}_1
  \partial_j \mathcal{T}_2 - \frac{D - 2}{4} \kappa_D^2 \mathcal{T}_1
  \mathcal{T}_2\,, \label{eqn:SDbranepressure}
\end{equation}
for Dirac branes.
Note that this reproduces~\eqref{eqn:boostInvariantPressure} in the special case of purely electric charges, see~\eqref{eqn:ttotau}. Likewise, for general branes
\begin{equation}
\mathcal{P}_{12} = 2 \pi t^{a b} Q_{1a} Q_{2b} - G^{i j} \mu_{1i} \mu_{1j} - \kappa_d^2 \biggl[\cT_1^{\mu \nu} \cT_{2\, \mu \nu} - \frac{1}{D-2} \cT_{1\; \mu}^{\mu} \cT_{2\; \nu}^{\nu} \biggr] \,. \label{eqn:SDgenbranepressure}
\end{equation}
by the same reasoning, c.f.~\eqref{eqn:branepressure}. 

\subsection{Black hole and black brane solutions}

Spherical symmetry allows only two components for $F_{d / 2}$, either along
the transverse $S^{d / 2}$ or radially with the remaining legs along the
brane. Thus,
\begin{equation}
  F^a = f_1^a (r) \omega_{d / 2} + f^a_2 (r) \ast \omega_{d / 2},
\end{equation}
for functions $f_1^a$ and $f_2^a$. The Bianchi identities imply $f_1^{a\prime}
(r) = 0$, where $f_2$ is fixed in terms of $f_1$ by the constraints.
Expressing the result in terms of the charge, we obtain:
\begin{equation}
  F^a = 2 \pi (t^{- 1} Q)^a  \frac{\ast \omega_{d / 2}}{V_{d / 2}} \mp 2 \pi
  (\eta^{- 1} Q)^a  \frac{\omega_{d / 2}}{V_{d / 2}} = F_1^a + (\Lambda^{- 1}
  \ast F_1)^a = (F_1 \mp \Lambda \ast F_1)^a,
\end{equation}
where
  $F_1^a = 2 \pi (t^{- 1} Q)^a  \frac{\ast \omega_{d / 2}}{V_{d / 2}}$
is the purely electric portion.

In terms of $F_1$, the stress tensor is
\begin{multline}
  T_{m n} = \frac{1}{4 \pi} t_{a b}  (F^a \circ F^b)_{m n} = \frac{1}{4 \pi}
  t_{a b}  (F^a_1 \circ F^b_1)_{m n} + \frac{1}{4 \pi}  (\Lambda^{\top} t
  \Lambda)_{a b}  (\ast F^a_1 \circ \ast F^b_1)_{m n} \\ = \frac{1}{2 \pi} t_{a
  b}  (F^a_1 \circ F^b_1)_{m n},
\end{multline}
where we make use of the fact that $(F^a_1 \circ \ast F^b_1)_{m n} = 0$ as
well as $\Lambda^{\top} t \Lambda = t$ and $\ast \omega \circ \ast \sigma =
\omega \circ \sigma$. Likewise,
\begin{equation}
  \frac{1}{4 \pi} \delta t_{a b} F^a \cdot F^b = \frac{1}{4 \pi} \delta t_{a
  b} F^a_1 \cdot F^b_1 + \frac{1}{4 \pi}  [\Lambda^{\top} \delta t \Lambda]_{a
  b} \ast F^a_1 \cdot \ast F^b_1 = \frac{1}{2 \pi} \delta t_{a b} F^a_1 \cdot
  F^b_1,
\end{equation}
where we use $\ast \omega \cdot \ast \sigma = - \omega \cdot \sigma$ and
  $\Lambda^{\top} \delta t \Lambda = \eta^{\top} t^{- 1} \delta t t^{- 1} \eta
  = - \eta^{\top} \delta t^{- 1} \eta = - \delta t$,
since $t = \eta^{\top} t^{- 1} \eta$.

Thus, the backreaction of $F^a$ on the metric and on the moduli is equivalent to
that of the electric flux $F_1^a$ with gauge kinetic matrix $\f_{a b} = \frac{1}{2 \pi} t_{a b}$,
and the relevant black hole potential is
\begin{equation}
  Q^2 (\phi) = 2 \pi t^{a b} (\phi) Q_a Q_b .
\end{equation}
Crucially, this is the same combination that appears in~\eqref{eqn:SDbranepressure}, \eqref{eqn:SDgenbranepressure}.
Apart from this replacement, the rest of the calculations in~\secref{sec:BH} are unchanged. 

\section{Example Solutions} \label{app:examples}

In this appendix, I discuss a few example black hole and black brane solutions
from the literature to further illustrate the discussion in
{\textsection}\ref{sec:BH}. For simplicity, I focus on solutions to the
Einstein-Maxwell-Dilaton effective action:
\begin{equation}
  S = \frac{1}{2 \kappa_d^2} \int d^d x \sqrt{- g}  \left( R_d - \frac{1}{2} 
  (\nabla \phi)^2 \right) - \frac{1}{2 e^2} \int d^d x \sqrt{- g} e^{- \alpha
  \phi} F_{p + 1}^2 . \label{eqn:dilAction}
\end{equation}
Thus, $G_{\phi \phi} (\phi) = \frac{1}{2 \kappa_d^2}$ and $f_{F F} (\phi) =
\frac{1}{e^2} e^{- \alpha \phi}$. By shifting the definition of $\phi$, we can set its
vacuum expectation value $\phi_{\infty}$ to zero at the expense of rescaling
$e^2$.

\subsection{Electrically charged solutions} \label{subsec:dilBHs}

Solutions with only electric or only magnetic charge have been studied
extensively in the literature \cite{gibbons:1982ih, myers:1986un, gibbons:1987ps, garfinkle:1990qj, horowitz:1991cd, lu:1993vt, duff:1993ye, duff:1996hp}, see also~\cite{Heidenreich:2015nta}. The two are related by Hodge duality,
which takes $p \rightarrow d - p - 2$, $\alpha \rightarrow - \alpha$, and $e^2
\rightarrow 4 \pi^2 / e^2$. Thus, both cases can be understood by studying
electrically charged solutions for general $p, \alpha$ and $e^2$.

The equations (\ref{eqn:psizEOM}), (\ref{eqn:phizEOM}) become
\begin{equation}
  d_z [f \dot{\psi}] = e^{2 \psi + \alpha \phi} \mathcal{Q}^2_0, \qquad d_z [f
  \dot{\phi}] = e^{2 \psi + \alpha \phi}  \frac{\alpha \mathcal{Q}^2_0}{\xi},
  \qquad \mathcal{Q}_0^2 = \frac{\xi e^2 Q^2 \kappa_d^2}{V_{d - p - 1}^2} ,
  \label{eqn:dilElEqns}
\end{equation}
where $\xi = \frac{p (d - p - 2)}{d - 2}$. A linear combination gives
  $d_z [f (\alpha \dot{\psi} - \xi \dot{\phi})] = 0$,
whose only regular solution satisfying 
$\psi_{\infty} = \phi_{\infty} = 0$ is $\alpha \psi - \xi \phi = 0$. Thus eliminating $\phi$ from (\ref{eqn:psiNoQ}), we find
  $\ddot{\psi} = \frac{1}{\xi \gamma}  \dot{\psi}^2 $ where $
  \gamma \df \left[ \xi + \frac{\alpha^2}{2} \right]^{- 1}$.
A general solution satisfying \ $\psi_{\infty} = 0$ is
\begin{equation}
  \psi (z) = - \xi \gamma \log \Bigl[ 1 + \frac{z}{z_0} \Bigr] \qquad
  \text{which implies} \qquad \phi (z) = - \alpha \gamma \log \Bigl[ 1 +
  \frac{z}{z_0} \Bigr],
\end{equation}
for some constant $z_0$. Substituting into (\ref{eqn:dilElEqns}), we obtain
  $\mathcal{Q}^2_0 = \frac{\xi \gamma}{z_0}  \left[ \frac{1}{z_0} +
  \frac{1}{z_h} \right]$.
A priori $z_0$ could have either sign, but if $z_0 < 0$ then regularity at the
horizon requires $z_0 > - z_h$ whereas $\mathcal{Q}_0^2 > 0$ requires $z_0 < -
z_h$, therefore $z_0 > 0$ by contradiction.

The thermodynamic properties of the solution can be read off using
(\ref{eqn:ADMmass}), (\ref{eqn:ADMtension}), (\ref{eqn:scalarcharge1}),
(\ref{eqn:hArea}), and (\ref{eqn:hGrav}):
\begin{subequations}
\begin{align}
  \mathcal{M} &= \frac{V_{d - p - 1}}{\kappa_d^2}  \biggl[ \frac{\gamma}{z_0} +
  \frac{d - p - 1}{d - p - 2}  \frac{1}{2 z_h} \biggr], & Q^2 &= \frac{V_{d - p - 1}^2}{e^2 \kappa_d^2}  \frac{\gamma}{z_0}  \biggl[
  \frac{1}{z_0} + \frac{1}{z_h} \biggr], \\
  \mathcal{T} &= \frac{V_{d - p - 1}}{\kappa_d^2}  \biggl[ \frac{\gamma}{z_0} +
  \frac{1}{d - p - 2}  \frac{1}{2 z_h} \biggr], & \mu_{\phi} &= \frac{V_{d - p - 1}}{\kappa_d^2}  \frac{\alpha \gamma}{2
  z_0}, \\
  s &= \frac{2 \pi}{\kappa_d^2} r_h^{d - p - 1} V_{d - p - 1}  \biggl[ 1 +
  \frac{z_h}{z_0} \biggr]^{\gamma}, &
  T &= \frac{d - p - 2}{4 \pi r_h}  \biggl[ 1 + \frac{z_h}{z_0} \biggr]^{-
  \gamma} . 
\end{align}
\end{subequations}
In the quasiextremal ($z_h = \infty$) case, $\mathcal{M}=
\frac{V_{d - p - 1}}{\kappa_d^2}  \frac{\gamma}{z_0}$ and $Q^2 = \frac{V_{d -
p - 1}^2}{e^2 \kappa_d^2}  \frac{\gamma}{z_0^2}$, so that $\kappa_d^2
\mathcal{M}^2 = \gamma e^2 Q^2$. More generally, $\kappa_d^2 \mathcal{M}^2
\geqslant \gamma e^2 Q^2$ with equality only at quasiextremality, so the quasiextremal solutions are extremal.

It is interesting to consider the behavior of the Hawking temperature as we
approach extremality. For $z_h \gg z_0$, we have:
\begin{equation}
  T \simeq \frac{(d - p - 2)^{\frac{d - p - 1}{d - p - 2}}}{4 \pi}
  z_0^{\gamma} z_h^{\frac{1}{d - p - 2} - \gamma} .
\end{equation}
Thus, either $T \rightarrow 0$ ($\gamma > \frac{1}{d - p - 2}$), $T
\rightarrow \text{constant}$ ($\gamma = \frac{1}{d - p - 2}$), or $T \rightarrow
\infty$ ($\gamma < \frac{1}{d - p - 2}$) as $z_h \to \infty$. All D$(p - 1)$ branes in $d=10$ string theory and its toroidal compactifications share the value
$\gamma = 1 / 2$. Thus, if $p < d - 4$ then $T \rightarrow 0$ at
extremality, whereas if $p = d - 4$ then $T \rightarrow \text{constant}$ at
extremality, and if $p > d - 4$ then $T \rightarrow \infty$ at extremality.
Therefore, cases with an apparently\footnote{Since the horizon is singular in
the quasiextremal limit, this behavior may be modified by derivative
corrections. I thank M. Montero, M. Reece and I. Valenzuela for discussions on this point.}
divergent Hawking temperature reside in the landscape of quantum gravities, e.g., for D6 branes
in ten-dimensional type IIA string theory.

By comparison, near extremality the entropy density behaves as
\begin{equation}
  s \simeq \frac{2 \pi V_{d - p - 1}}{\kappa_d^2}  (d - p - 2)^{- \frac{d - p
  - 1}{d - p - 2}} z_0^{- \gamma} z_h^{\gamma - \frac{d - p - 1}{d - p - 2}} .
\end{equation}
Notice that in general $\gamma \leqslant \frac{1}{\xi} = \frac{d - 2}{p (d - p
- 2)} \leqslant \frac{d - p - 1}{d - p - 2}$, so the entropy density either
goes to zero or to a constant in the extremal limit. In particular, the
inequalities are both saturated only when $\alpha = 0$ and $p = 1$, which is
the Reissner-Nordstr{\"o}m case, with extremal entropy:
\begin{equation}
  S_{\text{ext}} = 2 \pi \left( \frac{\kappa_d^2}{V_{d - 2}}
  \right)^{\frac{1}{d - 3}}  \left( \frac{M_{\text{ext}}}{d - 2}
  \right)^{\frac{d - 2}{d - 3}} .
\end{equation}
Otherwise, the entropy / entropy density goes to zero in the extremal limit.

\subsection{Dyonic solutions}

Dyonic solutions are possible for $d = 2 p + 2$.
Defining $\psi_e \df \psi + \frac{\xi}{\alpha} \phi$, $\psi_m \df \psi - \frac{\xi}{\alpha} \phi$,
(\ref{eqn:psizEOM}), (\ref{eqn:phizEOM}) become
\begin{equation}
  d_z [f \dot{\psi}_e] = 2\mathcal{Q}_e^2 e^{(1 + \nu) \psi_e + (1 - \nu)
  \psi_m}, \qquad d_z [f \dot{\psi}_m] = 2\mathcal{Q}_m^2 e^{(1 + \nu) \psi_m
  + (1 - \nu) \psi_e}, \label{eqn:dyonBHeqns}
\end{equation}
where $\nu \df \frac{\alpha^2}{2 \xi}$, $\mathcal{Q}_e^2 = \frac{\xi e^2 Q^2_e \kappa_d^2}{V_{d / 2}^2}$ and
$\mathcal{Q}_m^2 = \frac{\xi Q^2_m \kappa_d^2}{e^2 V_{d / 2}^2}$. Besides $\nu = 0$, explicit
solutions are known for $\nu = 1$~\cite{Lu:1995yn,duff:1996hp} and $\nu = 3$~\cite{Gibbons:1985ac}.

\subsubsection{$\boldsymbol{\nu = 1}$}
In this case, the equations (\ref{eqn:dyonBHeqns}) decouple into $d_z [f
\dot{\psi}_{e,m}] = 2\mathcal{Q}_{e,m}^2 e^{2 \psi_{e,m}}$, so that
\begin{equation}
  \psi_{e, m} = - \log \left[ 1 + \frac{z}{z_{e, m}} \right], \qquad
  \text{where} \qquad \frac{1}{z_{e, m}}  \left( \frac{1}{z_{e, m}} +
  \frac{1}{z_h} \right) = 2\mathcal{Q}_{e, m}^2 .
\end{equation}
The thermodynamic properties can be read off as before:
\begin{subequations}
\begin{align}
  \mathcal{M} &=  \frac{V_{d / 2}}{(d - 2) \kappa_d^2}  \biggl[ \frac{2}{z_e} +
  \frac{2}{z_m} + \frac{d}{2 z_h} \biggr], &
   Q^2_e &=  \frac{2 V_{d / 2}^2}{e^2 (d - 2) \kappa_d^2}  \frac{1}{z_e} 
  \biggl[ \frac{1}{z_e} + \frac{1}{z_h} \biggr], \\
  \mathcal{T} &=  \frac{V_{d / 2}}{(d - 2) \kappa_d^2}  \biggl[ \frac{2}{z_e} +
  \frac{2}{z_m} + \frac{1}{z_h} \biggr], &
  Q_m^2 &=  \frac{2 e^2 V_{d / 2}^2}{(d - 2) \kappa_d^2}  \frac{1}{z_m} 
  \biggl[ \frac{1}{z_m} + \frac{1}{z_h} \biggr],\\
  s &=  \frac{2 \pi}{\kappa_d^2} r_h^{d / 2} V_{d / 2}  \biggl[ 1 +
  \frac{z_h}{z_e} \biggr]^{\frac{2}{d - 2}}  \biggl[ 1 + \frac{z_h}{z_m}
  \biggr]^{\frac{2}{d - 2}}, &   \mu_{\phi} &=  \frac{V_{d / 2}}{\sqrt{2 (d - 2)} \kappa_d^2}  \biggl[
  \frac{1}{z_e} - \frac{1}{z_m} \biggr], \\
  T &=  \frac{d - 2}{8 \pi r_h}  \biggl[ 1 + \frac{z_h}{z_e} \biggr]^{-
  \frac{2}{d - 2}}  \biggl[ 1 + \frac{z_h}{z_m} \biggr]^{- \frac{2}{d - 2}} .
\end{align}
\end{subequations}
One can check that $\kappa_d \mathcal{M} \geqslant \sqrt{\frac{2}{d - 2}} 
\left[ | e Q_e | + \left| \frac{Q_m}{e} \right| \right]$, with equality only
in the quasiextremal case. For $r_h \ll r_{e, m} \df \left( \frac{d -
2}{2} z_{e, m} \right)^{- \frac{2}{d - 2}}$,
\begin{equation}
  s \simeq \frac{2 \pi V_{d / 2}}{\kappa_d^2} r_e r_m r_h^{\frac{d - 4}{2}},
  \qquad T \simeq \frac{d - 2}{8 \pi}  \frac{r_h}{r_e r_m},
\end{equation}
so $T \rightarrow 0$ in the extremal limit, whereas $s \rightarrow 0$ in this
limit for $d > 4$. In $d = 4$, the extremal entropy is finite
\begin{equation}
  S_{\text{ext}} = \frac{8 \pi^2}{\kappa_4^2} r_e r_m = \frac{1}{2}  | Q_e Q_m
  | . \label{eqn:extEntropy}
\end{equation}
Regardless of $d$, these solutions satisfy (\ref{eqn:attractorconds}) and have
an $\mathrm{AdS}_{d / 2} \times S^{d / 2}$ near horizon geometry of the form
(\ref{eqn:nearhorizon}).

\subsubsection{$\boldsymbol{\nu = 3}$}

This case arises naturally in Kaluza-Klein theory. Following~\cite{Gibbons:1985ac, Rasheed:1995zv}, we take the ansatz:
\begin{align}
  \psi_e &= - \frac{1}{2} \log \left[ 1 + \frac{2 z}{z_e} + \frac{z^2}{Z_e^2}
  \right], & \psi_m &= - \frac{1}{2} \log \left[ 1 + \frac{2 z}{z_m} +
  \frac{z^2}{Z_m^2} \right] .
\end{align}
Substituting into (\ref{eqn:dyonBHeqns}), we find
\begin{align}
  Z_e^2 &= \frac{z_e Z^2}{z_e + 2 z_h}, & Z_m^2 &= \frac{z_m Z^2}{z_m + 2
  z_h}, & \mathcal{Q}_e^2 &= \frac{(z_e + z_h) z_m}{2 z_e z_h Z_e^2},
  & \mathcal{Q}_m^2 &= \frac{(z_m + z_h) z_e}{2 z_m z_h Z_m^2},
\end{align}
where $Z^2 \df z_e z_h + z_m z_h + z_e z_m$. The thermodynamic properties
are now
\begin{subequations}
\begin{align}
  \mathcal{M} &= \frac{V_{d / 2}}{(d - 2) \kappa_d^2}  \biggl[ \frac{2}{z_e} +
  \frac{2}{z_m} + \frac{d}{2 z_h} \biggr], &
    Q_e^2 &= \frac{2 V_{d / 2}^2}{e^2  (d - 2) \kappa_d^2}  \frac{z_m}{Z_e^2} 
  \biggl[ \frac{1}{z_e} + \frac{1}{z_h} \biggr] , \\
  \mathcal{T} &= \frac{V_{d / 2}}{(d - 2) \kappa_d^2}  \biggl[ \frac{2}{z_e} +
  \frac{2}{z_m} + \frac{1}{z_h} \biggr], &
  Q_m^2 &= \frac{2 e^2 V_{d / 2}^2}{ (d - 2) \kappa_d^2}  \frac{z_e}{Z_m^2} 
  \biggl[ \frac{1}{z_m} + \frac{1}{z_h} \biggr] , \\
  s &= \frac{2 \pi}{\kappa_d^2} r_h^{d / 2} V_{d - 2}  \biggl[ \frac{(z_e +
  z_h) (z_m + z_h)}{Z_e Z_m} \biggr]^{\frac{2}{d - 2}} , &
  \mu_{\phi} &= \frac{\sqrt{3} V_{d / 2}}{\sqrt{2 (d - 2)} \kappa_d^2} 
  \biggl[ \frac{1}{z_e} - \frac{1}{z_m} \biggr] , \\
  T &= \frac{d - 2}{8 \pi r_h}  \biggl[ \frac{(z_e + z_h) (z_m + z_h)}{Z_e
  Z_m} \biggr]^{- \frac{2}{d - 2}} . 
\end{align}
\end{subequations}
In the quasiextremal limit, $Z_{e,m}^2 \rightarrow \frac{1}{2} z_{e,m}  (z_e + z_m)$, from which we obtain
\begin{equation}
  (\kappa_d \mathcal{M}_{\text{ext}})^{2 / 3} = \frac{1}{(d - 2)^{1 / 3}} 
  \Bigl( | e Q_e |^{2 / 3} + | Q_m/e |^{2 / 3} \Bigr) .
\end{equation}
One can show that all non-extremal solutions are heavier than this.\footnote{Note that the black hole region $ \bigl| \frac{e Q_e}{\kappa_d M} \bigr|^{2 / 3} + \bigl| \frac{Q_m}{e \kappa_d M} \bigr|^{2 / 3} \le (d-2)^{1/3}$ is not convex. This does not imply that the dyonic black holes are unstable against fragmentation into electric and magnetic constituents, because, due to the simultaneous presence of  electric and magnetic charge, angular momentum conservation plays a non-trivial role in the kinematics~\cite{Larsen:1999pp, Larsen:1999pu}.}

In the extremal limit, the entropy density and temperature behave similarly to the $\nu = 1$ case. Most notably, in $d = 4$,
\begin{equation}
  S_{\text{ext}} = \frac{8 \pi^2}{\kappa_4^2}  \frac{1}{Z_e Z_m} = \frac{1}{2}
  | Q_e Q_m | .
\end{equation}
The agreement with (\ref{eqn:extEntropy}) is due to the attractor mechanism.
Indeed, $S_{\text{ext}} = \frac{1}{2}  | Q_e Q_m |$ holds for all $\nu \neq
0$.

\bibliographystyle{JHEP}
\bibliography{refs}
\end{document}